\documentclass[11pt]{article}
\usepackage{amsmath}
\usepackage{amssymb}
\usepackage{amsthm}
\usepackage{bbm}
\usepackage{graphicx}
\usepackage{natbib}
\usepackage{booktabs}
\usepackage[left=25.4mm,right=25.4mm,top=30mm,bottom=25.4mm,a4paper]{geometry}
\usepackage{caption}
\usepackage{subcaption}
\usepackage{algorithm}
\usepackage[noend]{algpseudocode}
\usepackage{bm}

\theoremstyle{definition}
\newtheorem{theorem}{Theorem}

\newtheorem{remark}[theorem]{Remark}

\newtheorem{definition}{Definition}
\newtheorem{example}{Example}

\newcommand{\E}{\mathbb E}
\newcommand{\e}{\mathrm e}
\newcommand{\Q}{\mathbb Q}
\newcommand{\D}{\mathrm{d}}
\newcommand{\F}{\mathcal F}

\newcommand{\I}{\mathrm i}

\newcommand{\Var}{\mathrm{Var}}

\makeatletter
\def\BState{\State\hskip-\ALG@thistlm}
\makeatother

\begin{document}
\title{Computational method for probability distribution on recursive relationships in financial applications}
\author{Jong Jun Park\footnote{Department of Mathematical Sciences, KAIST, 291 Daehak-ro, Daejeon 34141, Republic of Korea}
	,\,\, Kyungsub Lee\footnote{Department of Statistics, Yeungnam University, Gyeongsan, Gyeongbuk 38541, Republic of Korea, Corresponding author, Email: ksublee@yu.ac.kr}
}
\date{}
\maketitle

\begin{abstract}
In quantitative finance, it is often necessary to analyze the distribution of the sum of specific functions of observed values at discrete points of an underlying process.
Examples include the probability density function, the hedging error, the Asian option, and statistical hypothesis testing.
We propose a method to calculate such a distribution, utilizing a recursive method, and examine it using various examples.
The results of the numerical experiment show that our proposed method has high accuracy.
\end{abstract}

\section{Introduction}

This paper introduces a recursive method to compute interesting quantities related to probability distributions in various financial applications.
The method is versatile, and hence, with slight modifications, it is easy to apply the basic framework to various applications.
More precisely, the method is based on a convolution-like formula, applied to compute the distribution of the sum of values of one-dimensional processes observed at discrete points.
Financial applications include numerical densities of asset price or volatility models, hedging error distributions, arithmetic Asian option prices, and statistical hypothesis tests.

Various kinds of stochastic processes are used in quantitative finance, such as the Cox-Ingersoll-Ross model (CIR), the constant elasticity of variance model (CEV), stochastic volatility, and GARCH models.
The probability distributions of the processes in financial models can be used for risk management, asset pricing, hedging analysis, parameter estimation, and statistical hypothesis testing.
In many cases, the closed form formulas for the density function of stochastic models are not known, and it is advantageous to develop a numerical procedure to compute the probability distributions or density functions.

When trading a financial option, the investor usually performs a hedging procedure to reduce risk.
In general, continuous models of asset price movements assume a continuous hedging process.
However, in practice, because continuous trading is not applicable, a discrete time hedging strategy is applied, and hence, the discrete time hedging error occurs even in the complete market model.
Many studies examine the discrete time hedging error in financial options.
\cite{sepp2012approximate} derived a numerically approximated distribution of the delta-hedging error based on the characteristic function for a jump diffusion model.
\cite{park2016distribution} computed the delta-hedging error based on a recursive method for a jump diffusion model, which is the same framework this study proposes.
This study extends this to a L\'{e}vy model and shows it is possible to easily adapt the method to not only delta-hedging processes but also other trading strategies, such as minimum variance hedging.

The arithmetic Asian option is a financial derivative whose payoff is the arithmetic average of the underlying asset prices observed at future times.
Asian options are safer with respect to the manipulation of underlying asset prices that may occur when they are close to maturity than European options and financial instruments suitable for less frequently traded assets \citep{musiela2006martingale}.
Since the closed-form formula is not available for the Asian option price, we study numerical approximation and simulation methods \citep{kemna1990pricing, Vecer2002}.
Our example is consistent with \cite{lee2014recursive}, which computes European option based Asian option prices; however, we directly apply risk-neutral probability density in this study.

We also examine an example of the statistical hypothesis test.
In general, a parametric statistical test depends on the probability distribution of a test statistic.
For typical sample mean tests, the test statistics are generally approximated by a t distribution; however, if the corresponding random variable is far from the normal distribution, it might undermine the accuracy of the test.
Therefore, a more exact distribution will be helpful in performing a more reliable test.
We provide an example of a skewness test where the recursive method is applied.
In general, financial asset return distributions are negatively skewed \citep{fama1965behavior,french1987expected,cont2001empirical}
and the third moment of financial asset distribution has been extensively studied \citep{KrausLitzenberger, HarveySiddique, Christoffersen2006, ChoeLee, lee2016probabilistic}.
Our example demonstrates a method to compute critical values and statistical power.

The rest of the paper is organized as follows:
Section~\ref{Sect:Basic} explains the basic recursive method and the process of applying the numerical procedure to compute the probability density functions.
Section~\ref{Sect:application} applies the proposed method to the examples.
Section~\ref{Sect:concl} concludes the paper.

\section{Basic method}\label{Sect:Basic}

\subsection{Derivation}
Let $X$ be a continuous stochastic process defined on the time horizon $[0, T]$ or
a discrete stochastic process defined on time indexes $0=t_0, t_1 , \cdots, t_N = T$.
If $X$ is continuous, we are particularly interested in the behaviors of $X_{t_i}$ for the discrete
observation times $0=t_0, t_1 , \cdots, t_N = T$.
When necessary, we can introduce a complete filtered probability space $(\Omega, \mathcal F, \mathbb P)$ over $[0, T]$, with filtration $\{\mathcal F_t \}_{t \in [0,T]}$.

In numerous financial applications, it is advantageous to examine the distribution of some finite summation of
$$Y = \sum_{i=1}^{N} h(X_{i-1}, X_i)$$
for some function $h$, where $X_i = X_{t_i}$.
A subsequent section will explain specific examples in financial practice, including a numerical probability density function, arithmetic Asian option price, the distribution of the realized variance, the hedging cost distribution, and statistical hypothesis testing.

To compute the distribution of $Y$, we propose a numerical scheme based on a recursive relationship.
One method to represent the conditional probability density function of $Y$ given $X_0 = x_0$ is based on the second derivative of the expectation of the European option payoff or the rectified unit linear function:
\begin{equation}
f_{Y|X_0}(y|x_0) = \frac{\D^2 \E[(Y -y)^+ | X_0 = x_0]}{\D y^2}.\label{Eq:basic}
\end{equation}
This approach is in line with the method introduced in \cite{breeden1978prices}, which derived the state price density function, which is similar to the risk-neutral density function, based on European option prices.
When the expectation is under a risk-neutral measure, the density function $f_{Y|X_0}$ is also considered to be the risk-neutral density function.
We are interested in both physical and risk-neutral probabilities.
This study examines the numerical method for computing the conditional probability density function $f_{Y|X_0}$ based on Eq.~\eqref{Eq:basic}.

To calculate
$$\E[(Y -y)^+] = \E \left[ \left( \sum_{i=1}^{N} h(X_{i-1}, X_i) -y \right)^+ \right],$$
consider the following relationship.
Define
$$ g_n(y|x_n) = \E \left[ \left. \left( \sum_{i=n+1}^{N}h(X_{i-1}, X_i ) - y \right)^+ \right| X_n = x_n \right]. $$
The above formula is similar to the $\F_{n}$-conditional expectation of the European option payoff, by regarding  
$\sum_{i=n+1}^{N}h(X_{i-1}, X_i )$ as an asset price and $y$ as a strike price.
Let $f_{X_{n+1}|X_n}(x_{n+1}| x_n)$ or simply $f(x_{n+1}| x_n)$, be the transition probability density function from $X_n = x_n$ to $X_{n+1} = x_{n+1}$.
Then, we derive the following successive relationships for $0 \leq n < N-1 $, as follows:
\begin{equation}
g_n(y | x_n) = \int_{\mathbb R} g_{n+1}(y-h(x_n ,x_{n+1})|x_{n+1}) f(x_{n+1}| x_n) \D x_{n+1} \label{Eq:relation1}
\end{equation}
and
\begin{align*}
g_{N-1}(y|x_{N-1}) &= \E [ (h(X_{N-1}, X_N) - y)^+ | X_{N-1} = x_{N-1}] \\
&= \int_{\mathbb R} (h(X_{N-1}, X_N) - y)^+ f(x_{N}| x_{N-1} ) \D x_{N}.
\end{align*}
Since
$$g_0(y|x_0) = \E [ (Y -y)^+ | X_0=x_0],$$
we have
$$ f_{Y|X_0}(y|x_0) = \frac{\partial^2 g_0(y|x_0)}{\partial y^2}.$$

Or, by differentiating both sides of Eq.~\eqref{Eq:relation1} for $y$, we can express
\begin{equation}
\bar F_n(y | x_n) = \int_{\mathbb R} \bar F_{n+1}(y-h(x_n ,x_{n+1})|x_{n+1}) f(x_{n+1}| x_n) \D x_{n+1} \label{Eq:relation2_bar}
\end{equation}
where $\bar F_n(y|x_n) = \frac{\partial g_n(y|x_n)}{\partial y} $,
and similarly,
\begin{align*}
\bar F_{N-1}(y|x_{N-1}) = \int_{\mathbb R} - {\mathbbm 1}_{\{y<h(x_{N-1}, x_N)\}} f(x_{N}| x_{N-1} ) \D x_{N}
\end{align*}
where $\mathbbm 1$ denotes the indicator function.
In the above equations, $\bar F_{n} = -1 + F_{n}$, where $F_{n}(y | x_n)$ is a conditional cumulative distribution function of $Y_n:=\sum_{i=n+1}^{N}h(X_{i-1}, X_i )$, given $X_n = x_n$.
Based on the discussion so far, we define the following.

\begin{definition}
We define
\begin{equation}
F_n(y | x_n) = \int_{\mathbb R} F_{n+1}(y-h(x_n ,x_{n+1})|x_{n+1}) f(x_{n+1}| x_n) \D x_{n+1} \label{Eq:relation2}
\end{equation}
and
\begin{equation}
F_{N-1}(y|x_{N-1}) = \int_{\mathbb R} \mathbbm 1_{ \{y\geq h(x_{N-1}, x_N)\}} f(x_{N}| x_{N-1} ) \D x_{N}. \label{Eq:relation2_end}
\end{equation}
In addition, by differentiating both sides of \eqref{Eq:relation2} for $y$, we obtain
\begin{equation}
f_n(y | x_n) = \int_{\mathbb R} f_{n+1}(y-h(x_n ,x_{n+1})|x_{n+1}) f(x_{n+1}| x_n) \D x_{n+1} \label{Eq:pdfrelation}
\end{equation}
and
\begin{equation}
f_{N-1}(y|x_{N-1}) =  \int_{\mathbb R}  \frac{\partial}{\partial y}\mathbbm 1_{ \{y\geq h(x_{N-1}, x_N)\}} f(x_{N}| x_{N-1} ) \D x_{N}  \label{Eq:pdfrelation2}
\end{equation}
where $f_n(y|x_n) = \frac{\partial F_n(y|x_n)}{\partial y} $, and the derivative is distributional.
\end{definition}


\begin{remark}
If $X$ is scale invariant, that is, $g_n(y|x_n)=x_n g_n\left(\frac{y}{x_n}|1\right)$, and let $\bar g_n(y) = g_n(y\big|1)$, we obtain
$$ g_{n+1}(y|x_{n+1}) = x_{n+1} g_{n+1} \left(\frac{y}{x_{n+1}}\Big|1 \right) = x_{n+1} \bar g_{n+1} \left(\frac{y}{x_{n+1}} \right)$$
and
\begin{align*}
g_n(y|x_n) &= \int_\mathbb{R} g_{n+1}(y-h(x_n ,x_{n+1})|x_{n+1}) f(x_{n+1}|x_n)  \D x_{n+1} \\
&= \int_\mathbb{R} x_{n+1} \bar g_{n+1} \left( \frac{y - h(x_n, x_{n+1})}{x_{n+1}} \right) f(x_{n+1}|x_n) \D x_{n+1}.
\end{align*}
By setting $x_n = 1$,
$$
\bar g_n(y) = \int_\mathbb{R} x_{n+1} \bar g_{n+1} \left( \frac{y - h(1, x_{n+1})}{x_{n+1}} \right) f(x_{n+1}|1) \D x_{n+1}
$$
and
$$ \bar g_{N-1}(y) = \E [ (h(1, X_N) - y)^+|X_{N-1}=1].$$
Alternatively, we can use the cumulative distribution function to express
$$
F_n(y) = \int_\mathbb{R} F_{n+1} \left( \frac{y - h(1, x_{n+1})}{x_{n+1}} \right) f(x_{n+1}|1) \D x_{n+1}
$$
and
$$ F_{N-1}(y) = \int_\mathbb{R} \mathbbm{1}_{\{ y \geq h(1,x_N)\}} f(x_N|1) \D x_N.$$
The above method applies to the one-dimensional function, and hence, the computational cost is much less.

One straightforward example is an arithmetic Asian option price under geometric Brownian motion.
In this case,
$$ 
g_0(y|x_n) = \E \left[ \left. \left( \sum_{i=1}^{N} X_i - y \right)^+ \right| X_0 = x_0 \right], $$
which can be interpreted as the Asian option price at time $0$, with underlying price $X_0 = x_0$, and strike price $y$.
Note that the price is equal to $x_0$ times the Asian option price with underlying price $X_0 = 1$ and strike price $y/x_0$.
In other words, $g_0(y|x_0)=x_0 g_0\left(\frac{y}{x_0}|1\right)$, and this is also applied to every $n$.

\end{remark}

\begin{remark}
	We obtain the intuitive forms by applying the Fourier transform to Eqs.~\eqref{Eq:pdfrelation}~and~\eqref{Eq:pdfrelation2}, and changing the order of integrals.
	By definition, the Fourier transform of the left-hand side of Eq.~\eqref{Eq:pdfrelation2} is the expectation of $\exp(-\I\nu Y_{N-1})$.
	\begin{eqnarray*}
		& &\int_{\mathbb{R}}\e^{-\I\nu y}f_{N-1}(y|x_{N-1})\D y\\
		&=&\int_{\mathbb{R}}\left[\int_{\mathbb R} \e^{-\I\nu y} \frac{\partial}{\partial y}\mathbbm 1_{ \{y\geq h(x_{N-1}, x_N)\}}\D y \right] f(x_{N}| x_{N-1} ) \D x_{N}\\
		&=&\int_{\mathbb{R}}\e^{-\I\nu h(x_{N-1},x_N)} f(x_{N}| x_{N-1} ) \D x_{N}\\
		&=&\mathbb{E}\left[\e^{-\I\nu Y_{N-1}}|x_{N-1}\right]
	\end{eqnarray*}
	Similarly, the Fourier transform of the left side of Eq.~\eqref{Eq:pdfrelation} is the expectation of $\exp(-\I\nu Y_{n})$.
	\begin{eqnarray*}
		& &\int_{\mathbb{R}}\e^{-\I\nu y}f_{n}(y|x_{n})\D y\\
		&=&\int_{\mathbb{R}}\left[\int_{\mathbb R} \e^{-\I\nu y} f_{n+1}(y-h(x_n ,x_{n+1})|x_{n+1}) \D y \right] f(x_{n+1}| x_n) \D x_{n+1}\\
		&=&\int_{\mathbb{R}}\e^{-\I\nu h(x_n,x_{n+1})}\left[\int_{\mathbb R} \e^{-\I\nu z} f_{n+1}(z)|x_{n+1}) \D z \right] f(x_{n+1}| x_n) \D x_{n+1}\\
		&=&\int_{\mathbb{R}}\mathbb{E}\left[\e^{-\I\nu (Y_{n+1}+h(x_n,x_{n+1}))}|x_{n+1} \right] f(x_{n+1}| x_n) \D x_{n+1}\\
		&=&\mathbb{E}\left[\e^{-\I\nu Y_{n}}|x_n\right]
	\end{eqnarray*}
	In the second equality, we substitute $z$ for $y-h(x_n ,x_{n+1})$.
\end{remark}

\subsection{Numerical procedure}\label{Sect:Numerical}

This subsection describes the numerical algorithm when applying the recursive method in computing a probability distribution.
There are specific considerations in applying the numerical method for every application; however, in this subsection, we examine the common concerns in applying the computational procedure.

\bigskip\noindent
{\it Selection of object function}

For the numerical procedure, we determine whether to use function $F_n$ or $f_n$, although this distinction has no significant effect on the results.
Both $F_n$ and $f_n$ have nice properties that stabilize the numerical procedure.
Since $F_n$ is theoretically bounded between 0 and 1, 
it can be easily corrected, even if $F_n$ is outside the bounded region, owing to numerical error.
The density function $f_n$ converges to 0 as $y$ goes $\infty$ or $-\infty$;
we can assume that integration with large $y$ in absolute value is almost zero.
This property of $f_n$ tends to make implementation easier when conducting numerical procedures; therefore, many examples in this study were based on $f_n$.
Meanwhile, in Subsection~\ref{Subsect:GARCH}, a singular point in the density function, such as in Dirac measure, makes it easier to use $F$ in the numerical procedure.

\bigskip\noindent
{\it Adaptive meshing}

For $F_n(y|x_n)$ or $f_n(y|x_n)$, the numerical domain is bounded by the region
$$ [x_{\textrm{min}}, x_{\textrm{max}}] \times [y_{\textrm{min}}, y_{\textrm{max}}]$$
where $x_{\textrm{min}}$ and $x_{\textrm{max}}$ are generally fixed,
whereas $y_{\textrm{min}}, y_{\textrm{max}}$ are usually dynamic throughout the iteration.
The adaptive change in the domain of $y$ during the numerical procedure is due to the change in the reasonable numerical support of the conditional distribution of $Y$ throughout the iteration.

For example, if the values of $y$ tend to increase as the numerical procedure proceeds, the numerical domain of $y$ changes accordingly.
Let us suppose our interest lies in the distribution of $Y = \sum_{i=1}^N X_i$.
The sufficient numerical domain for $\sum_{i=n}^{N} X_i$ is generally larger than $\sum_{i=n+1}^{N} X_i$.
Therefore, it is natural to expand the domain of $[y_{\textrm{min}}, y_{\textrm{max}}]$ for $F_n$, as $n$ proceeds from $N$ to 1.

If the numerical domain of $y$ is expanding, the number of intervals that divide the domain could become too large.
Instead of increasing the number of intervals over $y$,
we fix the total number of the discretized points over $[y_{\textrm{min}}, y_{\textrm{max}}]$
to prevent the grid size from becoming too large.
The dynamic allocation algorithm is straightforward.
Since $F_{n}$ or $f_{n}$ converges to 0 or 1, as $y$ approaches $\infty$ or $-\infty$ for all $x_n$,
with given threshold $\epsilon$, we increase $y_{\max}$ or decrease $y_{\min}$ as $F_{n}$ or $f_{n}$ reaches the convergence criteria; for example, $f(y_{\max}) < \epsilon$.
Since we fix the number of discretized points over $[y_{\textrm{min}}, y_{\textrm{max}}]$, say $M$,
the step size in $y$, $\Delta y  = (y_{\textrm{max}} - y_{\textrm{min}})/M$
this too changes, as $y_{\textrm{min}}, y_{\textrm{max}}$ change.

\bigskip\noindent
{\it Referencing previous function}

Another concern is referencing previous function values of $F_{n+1}$ or $f_{n+1}$ at step $n$,
as we compute
$$
F_n(y | x_n) = \int_{\mathbb R} F_{n+1}(y-h(x_n ,x_{n+1})|x_{n+1}) f(x_{n+1}| x_n) \D x_{n+1}.
$$
When query point $(y-h(x_n ,x_{n+1}), x_{n+1})$ falls within the grid  $[y_{\textrm{min}}, y_{\textrm{max}}] \times [x_{\textrm{min}}, x_{\textrm{max}}]$ defined at time $t_{n+1}$,
we simply retrieve values such as $F_{n+1}(y-h(x_n ,x_{n+1})|x_{n+1})$ or $ f_{n+1}(y-h(x_n ,x_{n+1})|x_{n+1})$ using interpolations, such as
linear or piecewise cubic Hermite interpolations.
This works well even using the nearest value.

When $y-h(x_n ,x_{n+1})$ is outside $[y_{\textrm{min}}, y_{\textrm{max}}]$,
we assign the extrapolated value depending on the function, $F_{n+1}$ or $f_{n+1}$.
Since $F_{n+1}$ is a cumulative distribution function, $F_{n+1}(y| \cdot)$ approaches 1 when $y$ approaches $\infty$ and $F_{n+1}(y| \cdot)$ approaches $0$ when $y$ approaches $-\infty$.
Therefore, it is natural to assign 
$$F_{n+1}(y-h(x_n ,x_{n+1})| \cdot) = 1$$
when $y-h(x_n ,x_{n+1}) > y_{\textrm{max}}$
and 
$$F_{n+1}(y-h(x_n ,x_{n+1})| \cdot) = 0$$
when $y-h(x_n ,x_{n+1}) < y_{\textrm{min}}$.
Similarly, for $f_{n+1}$, we assign $f_{n+1}(y-h(x_n ,x_{n+1})| \cdot) = 0$ when $y-h(x_n ,x_{n+1}) > y_{\textrm{max}}$ or $y-h(x_n ,x_{n+1}) < y_{\textrm{min}}$.

\begin{figure}
\centering
\includegraphics[width=0.4\textwidth]{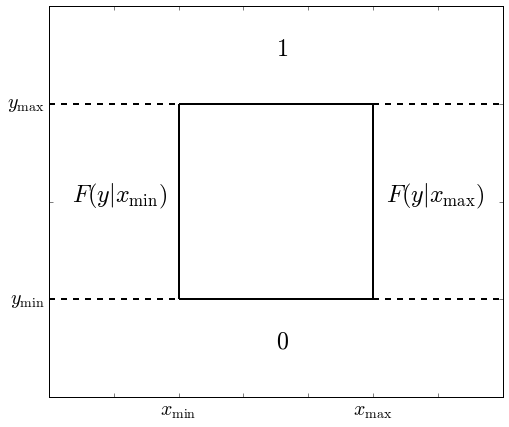}
\quad\quad
\includegraphics[width=0.4\textwidth]{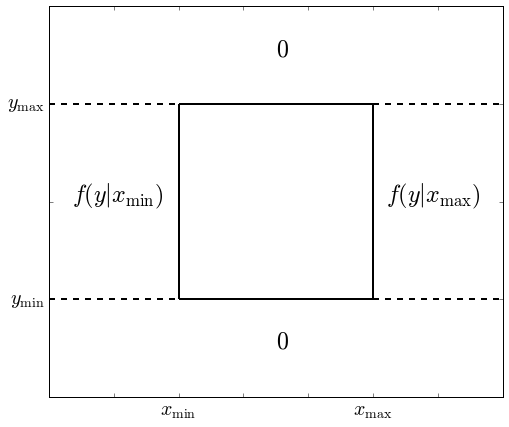}
\caption{Reference criteria for $F_n$ (left) and $f_n$ (right) outside the numerical domains}\label{Fig:criteria}
\end{figure}

When $x_{n+1}$ is outside $[x_{\textrm{min}}, x_{\textrm{max}}]$ and $y-h(x_n ,x_{n+1}) \in [y_{\textrm{min}}, y_{\textrm{max}}]$,
we assign 
$$F_{n+1}(y-h(x_n ,x_{n+1})|x_{n+1}) = F_{n+1}(y-h(x_n ,x_{n+1})|x_{\textrm{max}})$$
when $x_{n+1} > x_{\textrm{max}}$
and 
$$F_{n+1}(y-h(x_n ,x_{n+1})|x_{n+1}) = F_{n+1}(y-h(x_n ,x_{n+1})|x_{\textrm{min}})$$
when $x_{n+1} < x_{\textrm{min}}$.
In this case, $x_{n+1}$ is far from $x_{n}$ and the transition probability $f(x_{n+1}|x_{n})$ is relatively small,
and hence, $F_{n+1}(y-h(x_n ,x_{n+1})|x_{n+1}) f(x_{n+1}|x_{n})$ is close to zero.
Therefore, error due to discrepancy between the true value of $F_{n+1}(y-h(x_n ,x_{n+1})|x_{n+1})$ and its approximation is small, relative to the overall integration.
We summarize this explanation in Figure~\ref{Fig:criteria}
and present the algorithm in Algorithm~\ref{method}\footnote{For an example of Matlab code, see {https://github.com/ksublee/Recursive\_method}}.

\begin{algorithm}[h]
	\caption{Recursive method with distribution function}\label{method}
	\begin{algorithmic}

		\State \texttt{Initial setting for $\bm{F}_{N-1}$ over a grid  $[x_{\textrm{min}}, x_{\textrm{max}}] \times [y_{\textrm{min}}, y_{\textrm{max}}]$ using} Eq.~\eqref{Eq:relation2_end}
		\For{ $n \in \{ N-2, \cdots,  1 \}$  } 
		
		\For{ $x \in \{ x_{\min}, \cdots, x_{\max}  \}$ } \label{line:start}
		\State \texttt{Set a vector $\bm{\chi}$ for $X_{n+1}$ around $X_n = x$}
		\State $\bm{\rho} \gets $\texttt{ pdf vector of $X_{n+1}$ over $\bm{\chi}$ with $X_n = x$ }
		\For{ $y \in \{ y_{\min}, \cdots, y_{\max}  \}$ }
		
		\State $\bm{\psi} \gets  \bm{F}_{n+1}[\bm{\chi}, y-h(x, \bm{\chi})]$ \Comment{Use reference rule in Figure~\ref{Fig:criteria}}
		\State $\bm{F}_n[x, y] \gets \int \bm{\rho} \bm{\psi} $ \Comment{Numerical integration} \label{line:end}
		
		\EndFor
		\EndFor
		
		\While {\texttt{$\bm{F_n}[,y_{\min}] \nsim 0 $ or $\bm{F_n}[,y_{\max}] \nsim 1$ }} 
		
		\State \texttt{Extend grid by $y_{\min} \gets y_{\min} - \Delta y$ or $y_{\max} \gets y_{\max} + \Delta y$}
		\State \texttt{Repeat lines from \ref{line:start} to \ref{line:end} for new points}
		
		\EndWhile
		
		\EndFor
		
	\end{algorithmic}
\end{algorithm}

\subsection{Computational cost}

The computational cost depends on the number of time steps and the size of grid $ [x_{\textrm{min}}, x_{\textrm{max}}] \times [y_{\textrm{min}}, y_{\textrm{max}}]$. 
To compute $f_n (y | x)$, we perform numerical integration for every point $(x,y)$ in the grid.
Let the numbers of $x$ and $y$ in the grid be $N_x$ and $N_y$, respectively,
and the number of steps for numerical integration in Eqs~\eqref{Eq:relation2} or \eqref{Eq:pdfrelation} be $N_z$.
Then, the computational complexity of each time step is proportional to $N_x N_y N_z$.

This is similar to an alternative method such as the Fourier transform described in Remark~\ref{remark:Fourier}.
More precisely, let $\xi_n  = \E[\e^{-\I \nu Y} | \F_n]$.
Then, the equation corresponding to the recursive relationship is
$$\xi_{n-1}(x_{n-1}, \nu)  = \int_{\mathbb R} \xi_n(x_n, \nu) \e^{-\I \nu h(x_n, x_{n-1})} f(x_n, x_{n-1}) \D x_n $$,
and hence, we perform numerical integration with respect to $x_{n}$ on every point of $(x_{n-1}, \nu)$ over
a grid $ [x_{\textrm{min}}, x_{\textrm{max}}] \times [\nu_{\textrm{min}}, \nu_{\textrm{max}}]$, for every time step.
Therefore, essentially, the Fourier transform and the recursive methods have the same time complexity.
One advantage of our method over the Fourier method is that we do not have to apply Fourier transform in the final step to retrieve the distribution or density function.

\section{Application}\label{Sect:application}

In this section, we apply our proposed method to several examples.
As we introduce various distinct examples, please note that each subsection uses different notations.

\subsection{Numerical density for diffusion models}

This subsection demonstrates the computation of probability densities, or likelihood functions of various diffusion models numerically based on the recursive method.

\subsubsection{CIR model}

Consider a square-root process $X$, also known as the Cox-Ingersoll-Ross model \citep{cox1985theory}, defined by:
$$ \D X_t = \kappa(\theta - X_t) \D t + \gamma \sqrt{X_t} \D W_t .$$
The density function of the transition probability from $X_0 = x_0$ to $X_t = x$ of the square root process is given by
$$f(x|x_0)= c\exp(-u-c(x+x_0)) \left\{ \frac{c(x+x_0)}{u} \right\}^{q/2}I_q\left(2\sqrt{uc(x+x_0)}\right)$$
where
$$ c = \frac{2\kappa}{(1-\exp(-\kappa t))\gamma^2}, \quad u = c x_0 \exp(-\kappa t), \quad q = 2\kappa\theta/\gamma^2 - 1 $$
and
$I_q$ denotes the modified Bessel function of the first kind of order $q$.

Although the closed-form solution of the density function is available, for illustrative purposes, we examine the approximation method based on recursive relation and discretization to compute the probability density function of $X_N$ for some $t_N$.
The approximate distribution of $\Delta X_n = X_n - X_{n-1}$  with $\Delta t = t_n - t_{n-1}$ by normal distribution in typical Monte Carlo simulations is as follows:
$$ \Delta X_n \sim N\left( \kappa(\theta - X_{n-1} )\Delta t, \gamma \sqrt{X_{n-1}} \right) $$
Let
$$ h(x_{i-1}, x_i) = x_i - x_{i-1},$$
then
$$Y = \sum_{i=1}^{N} h(X_{i-1}, X_i) = X_N - X_0$$
and the recursive method can be applied to compute the distribution of $Y$, that is, $X_N-X_0$.
For the integrand of the recursive method, we use the conditional probability density function $f_n(y|x_n)$.
We obtain similar results using the conditional distribution function $F_n(y|x_n)$.

\begin{figure}
\centering
\includegraphics[width=0.4\textwidth]{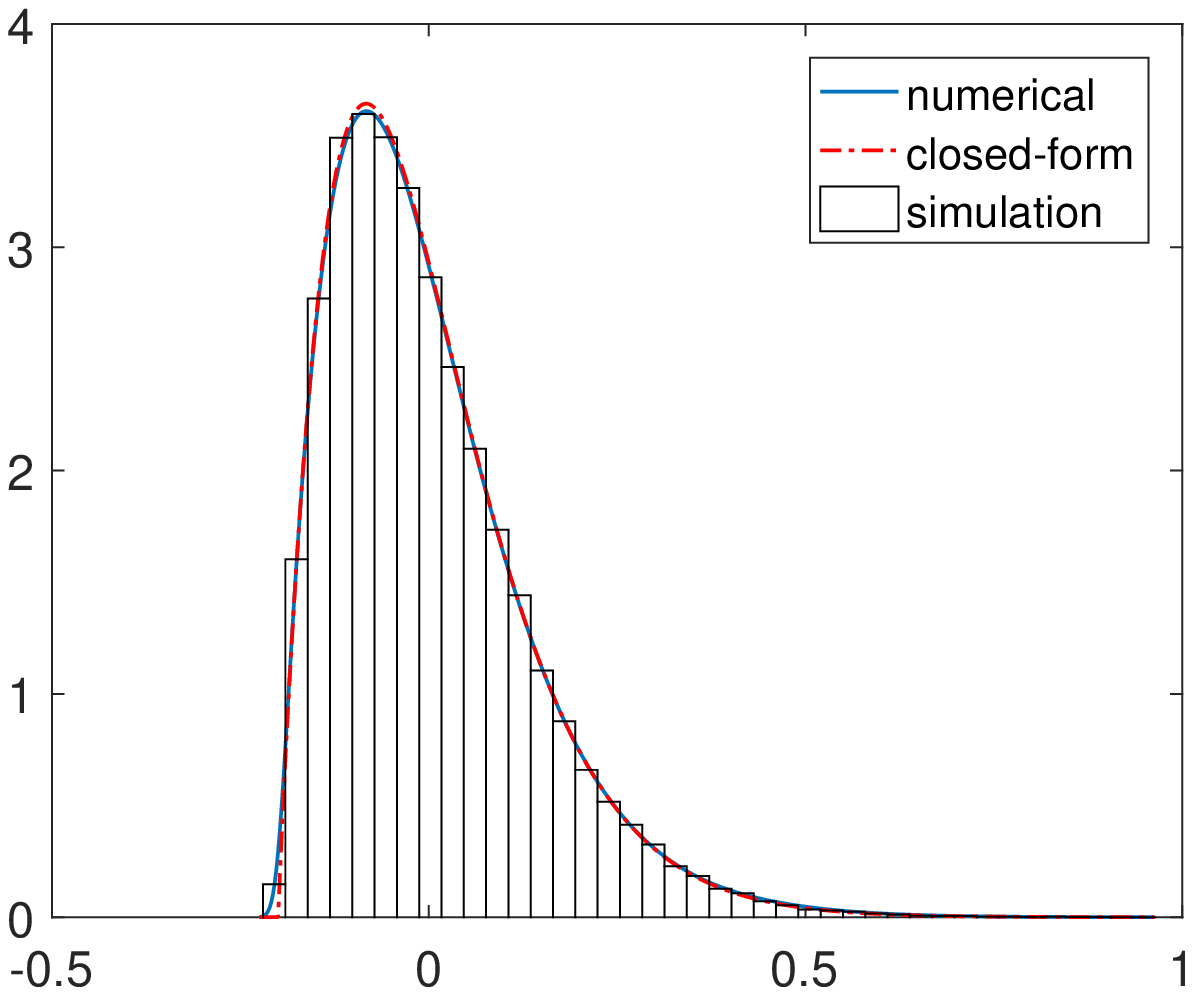}
\quad
\includegraphics[width=0.42\textwidth]{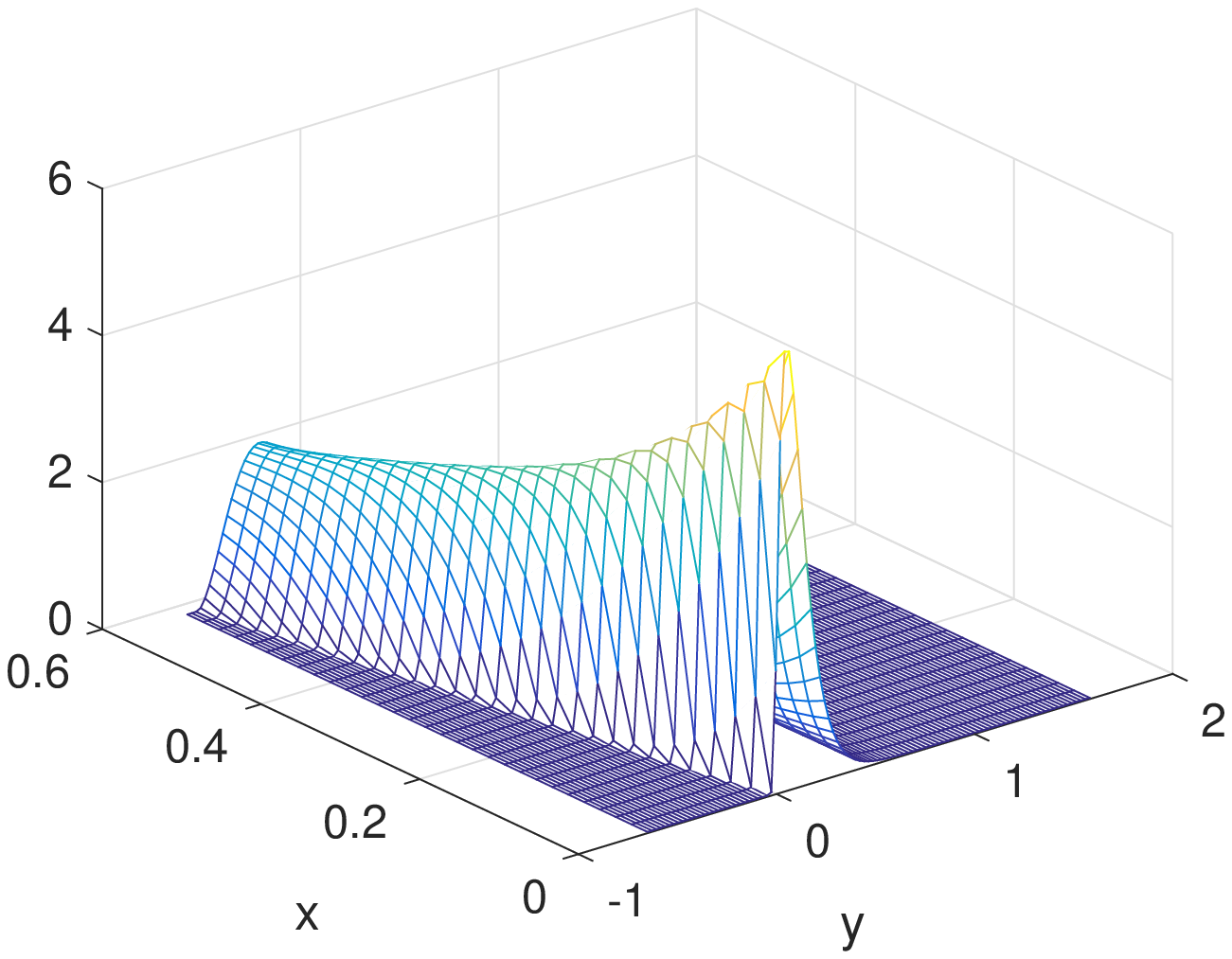}
\caption{Probability density function of the CIR model (left) and $f_{Y|X_0}(y|x_0)$ (right)}\label{Fig:SQRT}
\end{figure}

\begin{figure}
\centering
\includegraphics[width=0.5\textwidth]{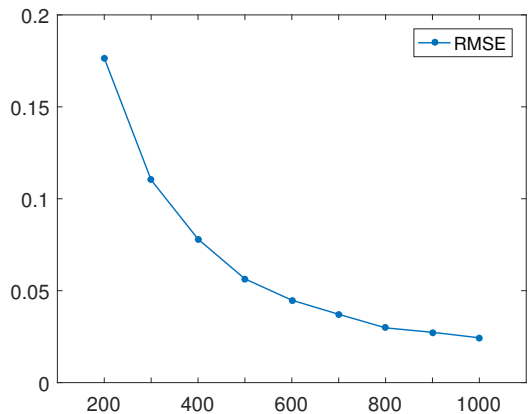}
\caption{Global error for probability density function of the CIR model}\label{Fig:RMSE}
\end{figure}

Figure~\ref{Fig:SQRT} compares the numerically computed density function, closed-form formula, and simulation results.
The parameter settings for the square root process are $\kappa=11, \theta = 0.2, \gamma = 1.5$
and for the numerical procedure
$\Delta t = 1/1250, N=100, [x_{\min} , x_{\max} ] = [0, 0.6]$ with $\Delta x = 0.002$;
The number of intervals for the $y$ domain are 1,000.
Time is annualized, and hence, $t_N = 20$ days.
On the left side of Figure~\ref{Fig:SQRT}, the numerically computed density function based on the recursive method is very close to the closed-form formula and simulation histogram.
On the right side of Figure~\ref{Fig:SQRT}, the conditional density $f_{Y|X_0}$ is plotted as a function of $X_0 = x_0$ and $Y = X_N - X_0 =y$.
The global error is decreasing with the increasing number of intervals for the $y$-axis
as plotted in Fig~\ref{Fig:RMSE}.

\subsubsection{CEV model}

The constant elasticity of variance model (CEV) identifies the leverage effect of volatility being negatively correlated with asset prices \citep{Cox1976}
with the following formula in the stock price process:
$$ \D X_t = \mu X_t \D t + \sigma X_t^\gamma \D W_t.$$
The closed-form formula of the stock price distribution is not known
and it is worthwhile to compute the density function with the numerical method.
The conditional probability density function is computed using the same method as the square root process in the previous subsection.

\begin{figure}
\centering
\includegraphics[width=0.4\textwidth]{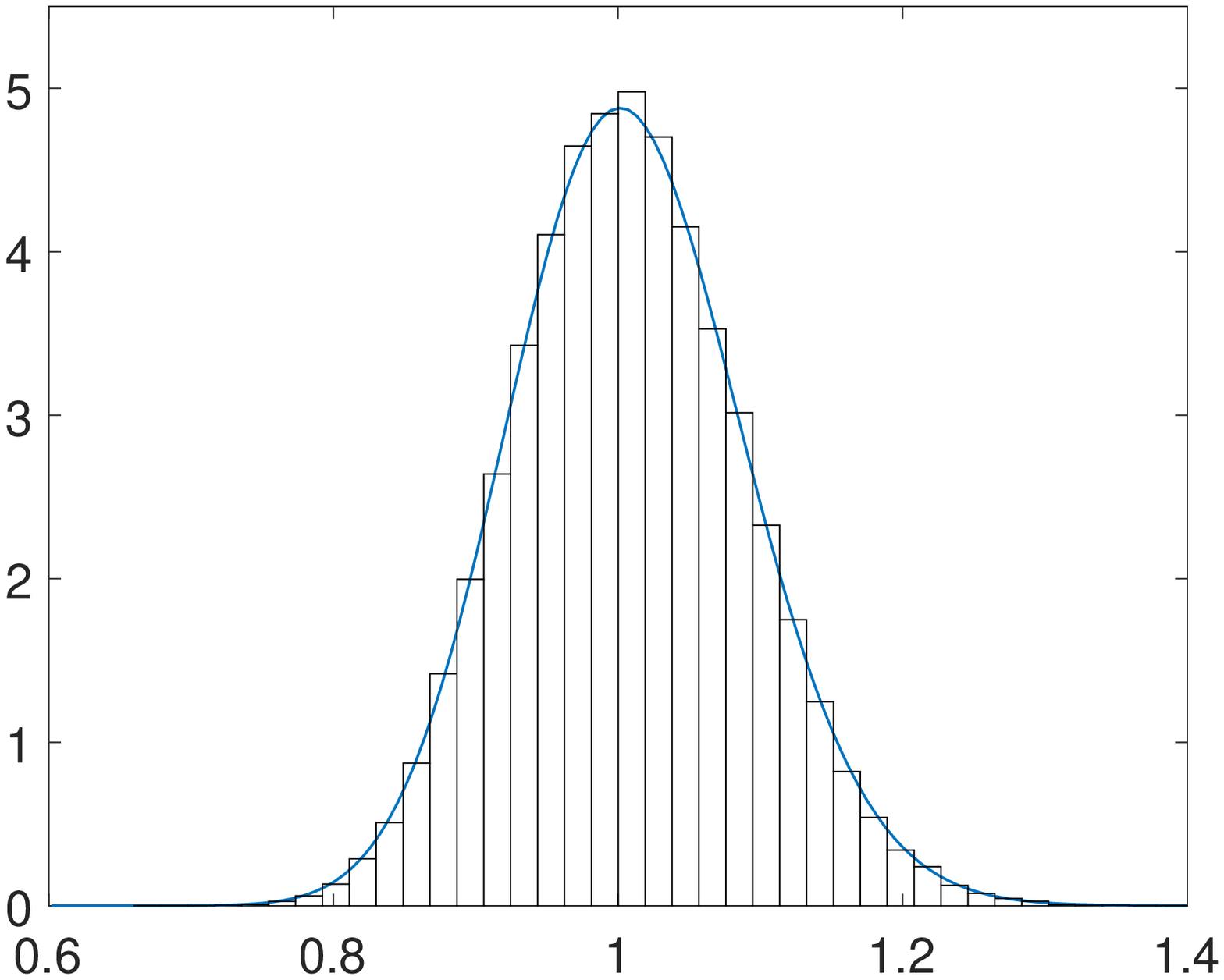}
\quad
\includegraphics[width=0.42\textwidth]{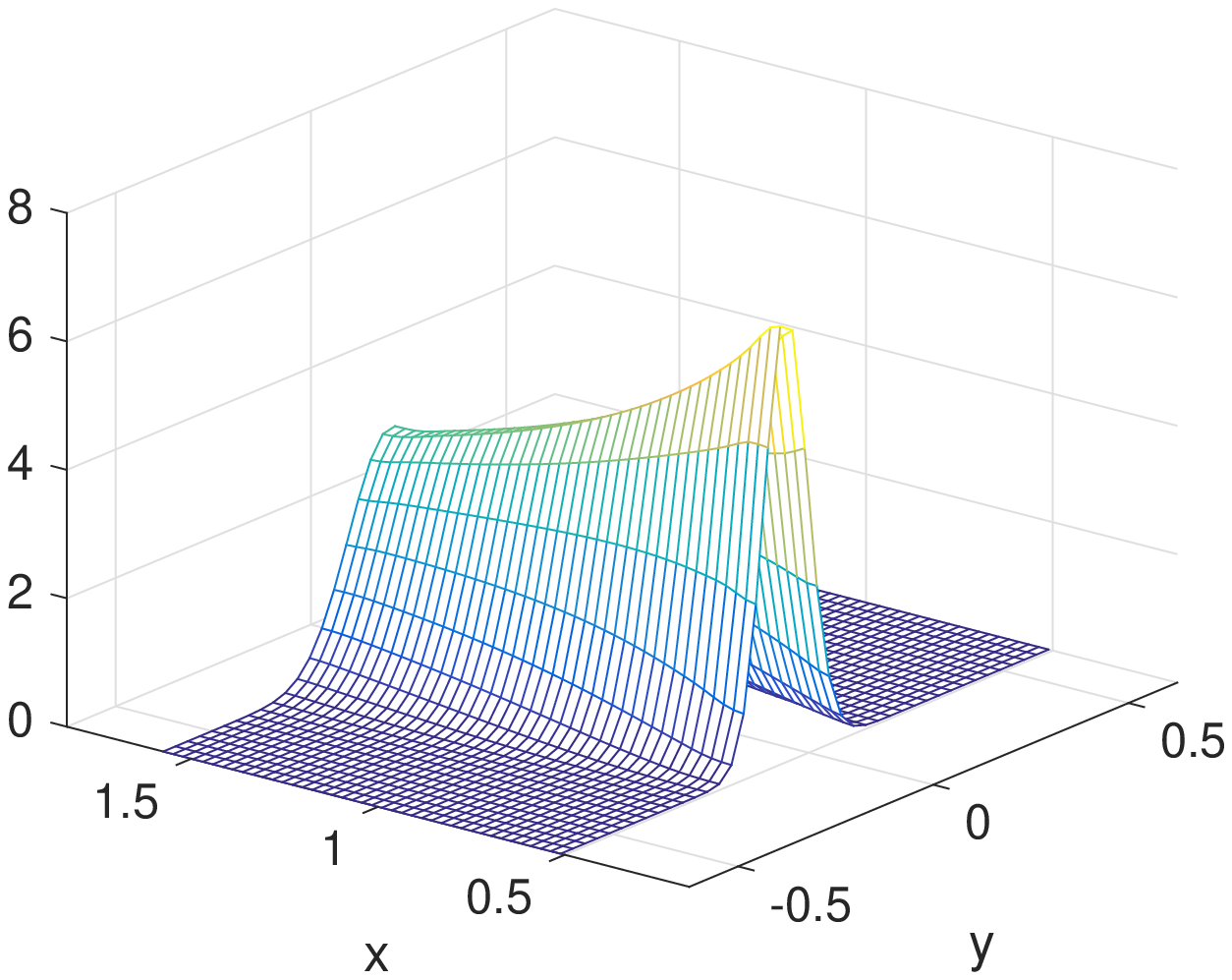}
\caption{Probability density function of the CEV model (left) and $f_{Y|X_0}(y|x_0)$ (right) }\label{Fig:CEV}
\end{figure}

The conditional probability density functions of $Y$, $f_{Y|X_0}(y|x_0)$ with various $x_0$ (right)
and the density function of $X_t$ with $X_0 = 1$ is presented in Figure~\ref{Fig:CEV}.
Comparing the numerical probability density function and the simulation histogram shows that the recursive method generates a more precise density function.
The parameter setting for the CEV model example is $\mu=0.05, \sigma = 0.2, \gamma = 0.7$.
For the numerical procedure,
$\Delta t = 1/1250, N=200, [x_{\min} , x_{\max} ] = [0.5, 1.5]$ with $\Delta x = 0.005$,
the number of intervals on the $y$-axis are 200 and the tolerance for the dynamic allocation is $10^{-8}$.

\subsubsection{Stochastic volatility model and integrated variance}

This subsection computes the probability density function based on the numerical likelihood method
for various stochastic volatility models.
Consider a stochastic volatility model such that
$$ \D V_t = \kappa V^a (\theta - V_t) \D t + \gamma V^b \D W_t $$
with parameters of $a \in \{0, 1 \}$ and $b \in \{1/2, 1, 3/2\}$.
This classification is from \cite{Christoffersen2010}.
When $a=0$ and $b=1/2$, the square root process is used for stochastic volatility, as in \cite{Heston1993}.

\begin{figure}
\centering
\begin{subfigure}[b]{0.45\textwidth}
\includegraphics[width=\textwidth]{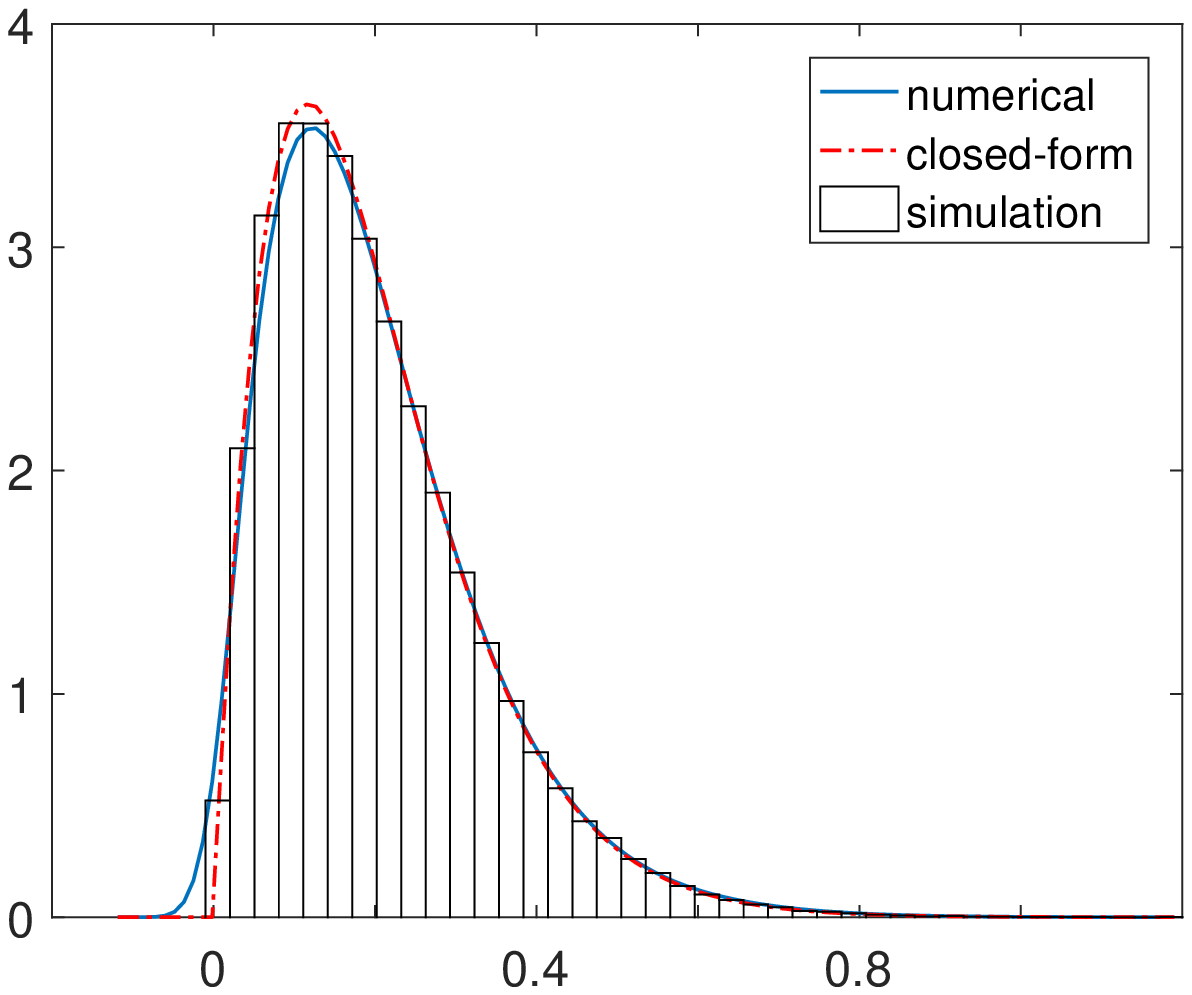}
\caption{$a = 0, b=1/2$}
\end{subfigure}
\begin{subfigure}[b]{0.45\textwidth}
\includegraphics[width=\textwidth]{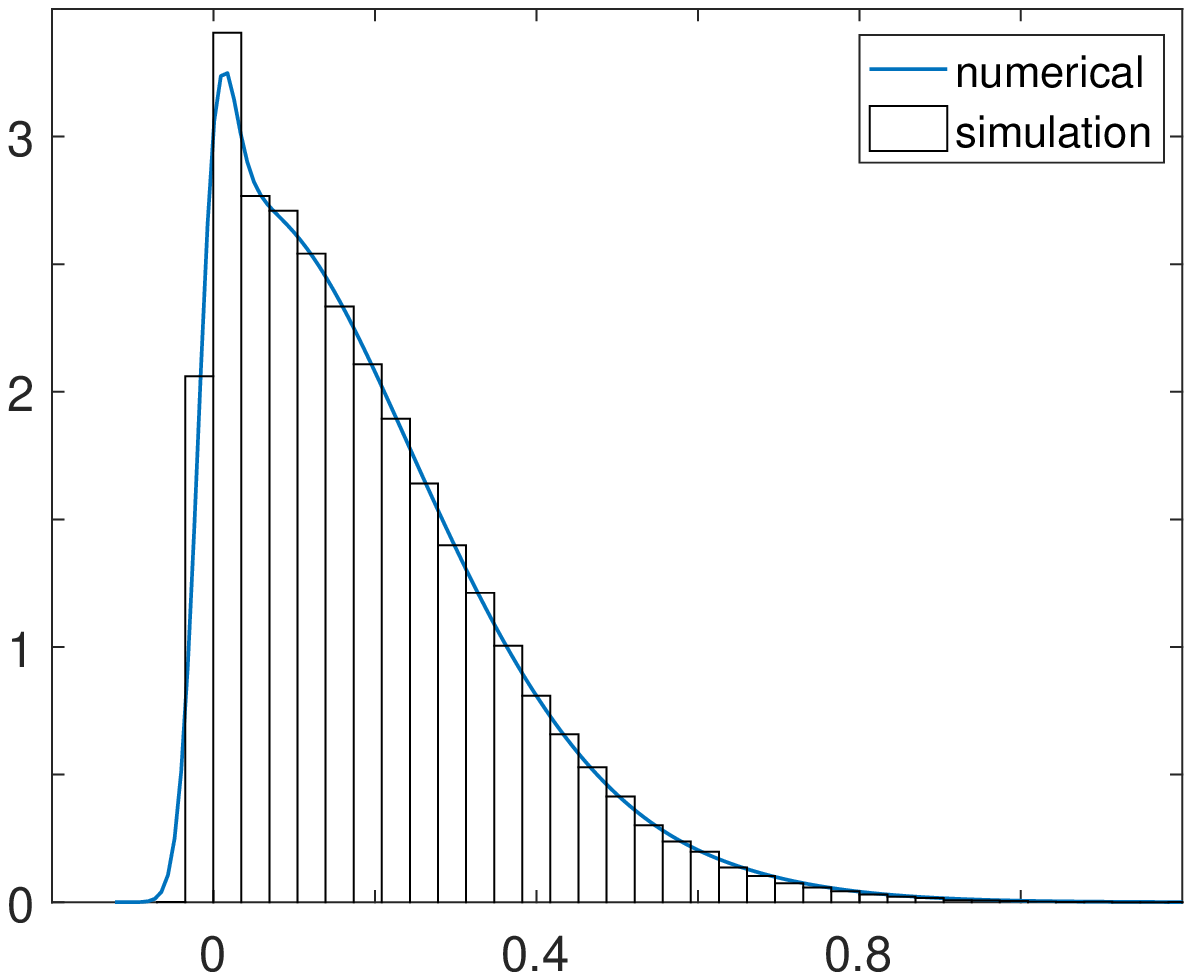}
\caption{$a = 1, b=1/2$}
\end{subfigure}
\\
\begin{subfigure}[b]{0.45\textwidth}
\includegraphics[width=\textwidth]{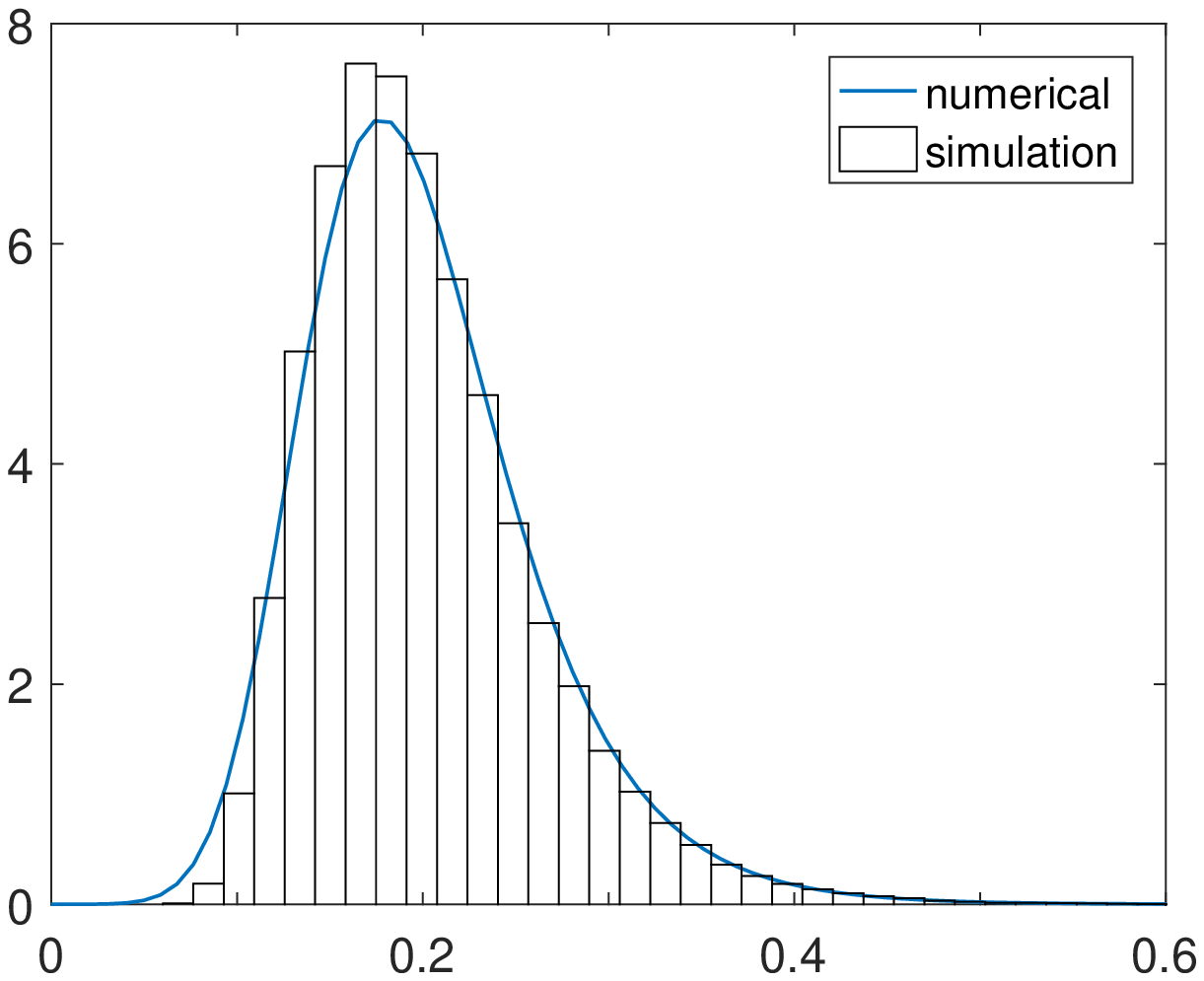}
\caption{$a = 0, b=1$}
\end{subfigure}
\begin{subfigure}[b]{0.45\textwidth}
\includegraphics[width=\textwidth]{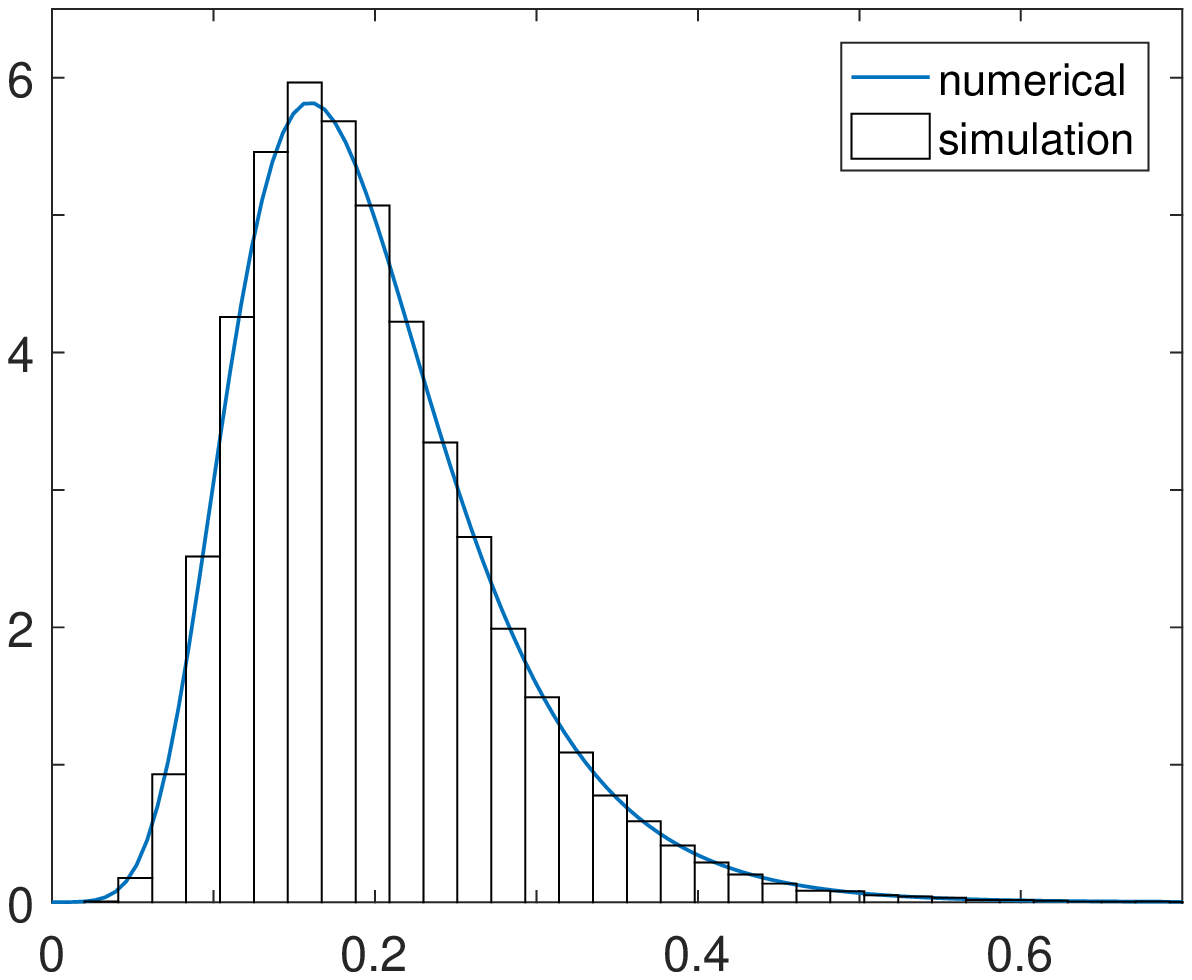}
\caption{$a = 1, b=1$}
\end{subfigure}
\\
\begin{subfigure}[b]{0.45\textwidth}
\includegraphics[width=\textwidth]{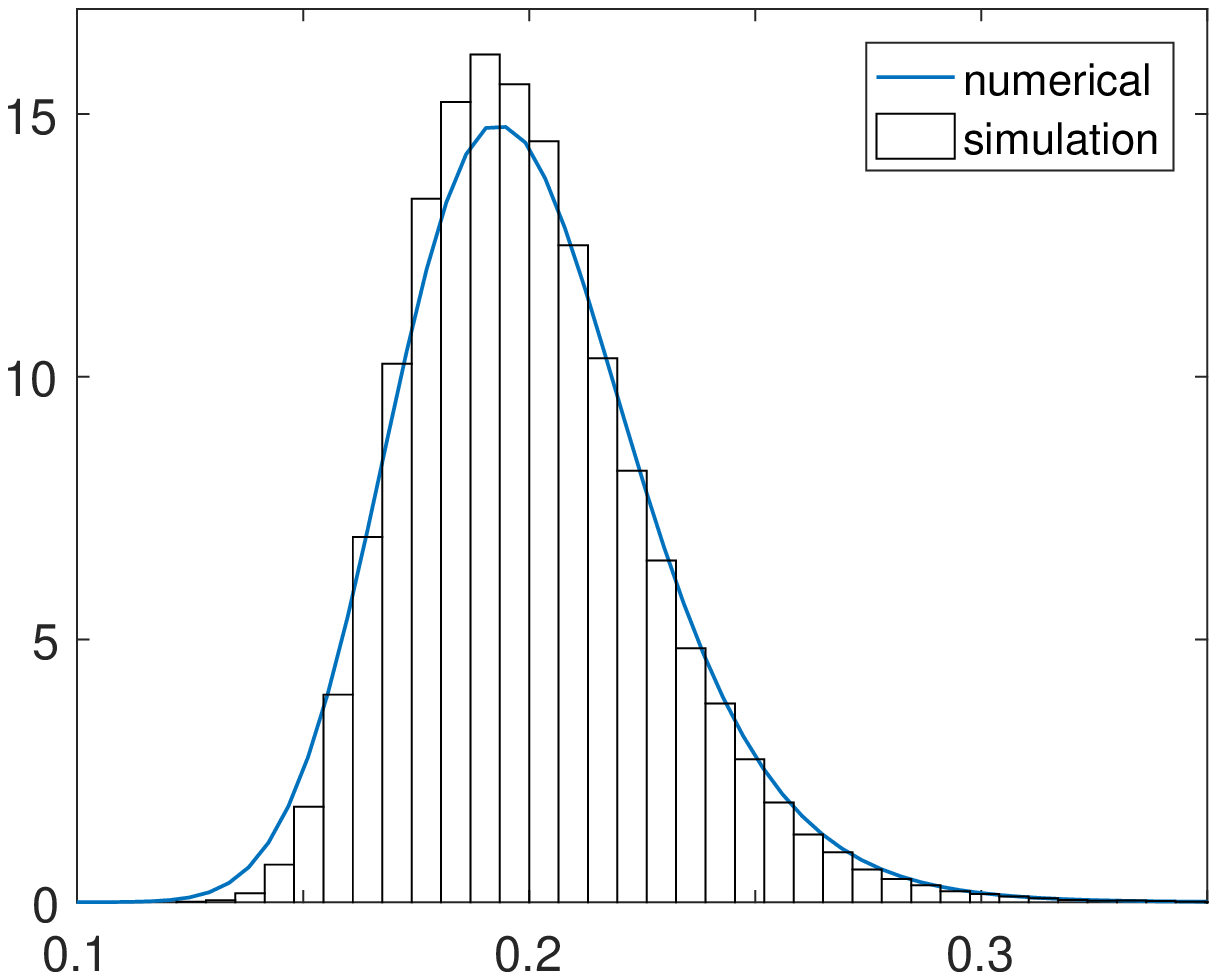}
\caption{$a = 0, b=3/2$}
\end{subfigure}
\begin{subfigure}[b]{0.45\textwidth}
\includegraphics[width=\textwidth]{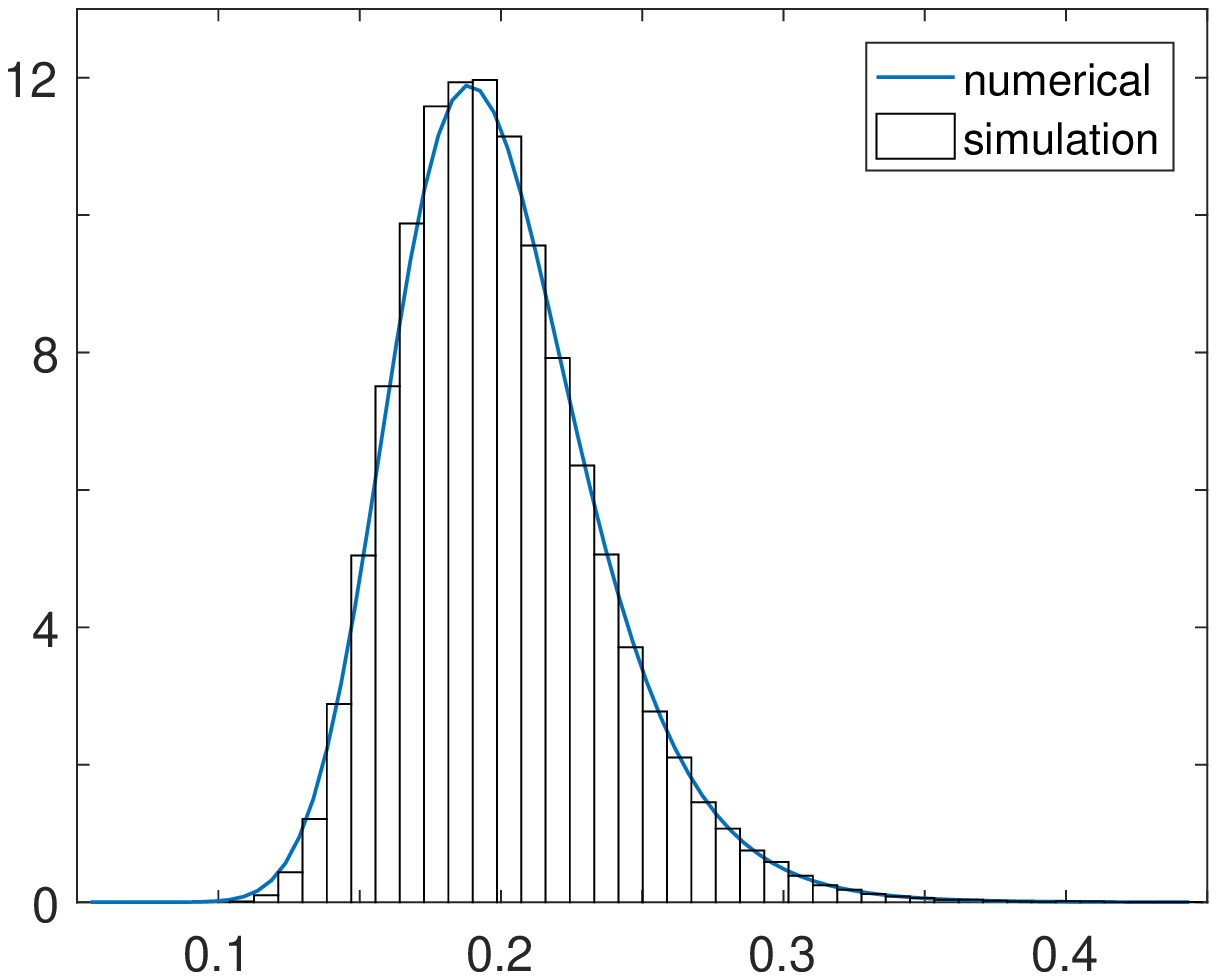}
\caption{$a = 1, b=3/2$}
\end{subfigure}
\caption{Probability density function of $Y = V_T - V_0$ with various stochastic volatility models}\label{Fig:SV}
\end{figure}

Other than $a$ and $b$, we fix the parameter setting $\kappa = 11, \theta = 0.2, \gamma = 0.8, V_0 = 0.2$,
and for the numerical procedure, $\Delta t = 1/1250,  N = 100$.
Figure~\ref{Fig:SV} presents the numerically computed various probability density functions.
For stochastic volatility models, the numerical probability density functions are close to the simulation histograms.

The integrated variance is defined by
$$IV_t = \frac{1}{t}\int_0^t V_s \D s.$$
It is known that the realized variances in the stochastic volatility models
converge to the integrated variances \citep{barndorff2002econometric}.
However, the closed-form formulas of the unconditional distribution for integrated variances of stochastic volatility
are generally not known.
For discussion on special cases of affine models, see \cite{broadie2006exact}.
Since the variance process is unobservable, sometimes the integrated variance is more interesting, as it can be approximated by the quadratic process of underlying return.

\begin{figure}
\centering
\begin{subfigure}[b]{0.45\textwidth}
\includegraphics[width=\textwidth]{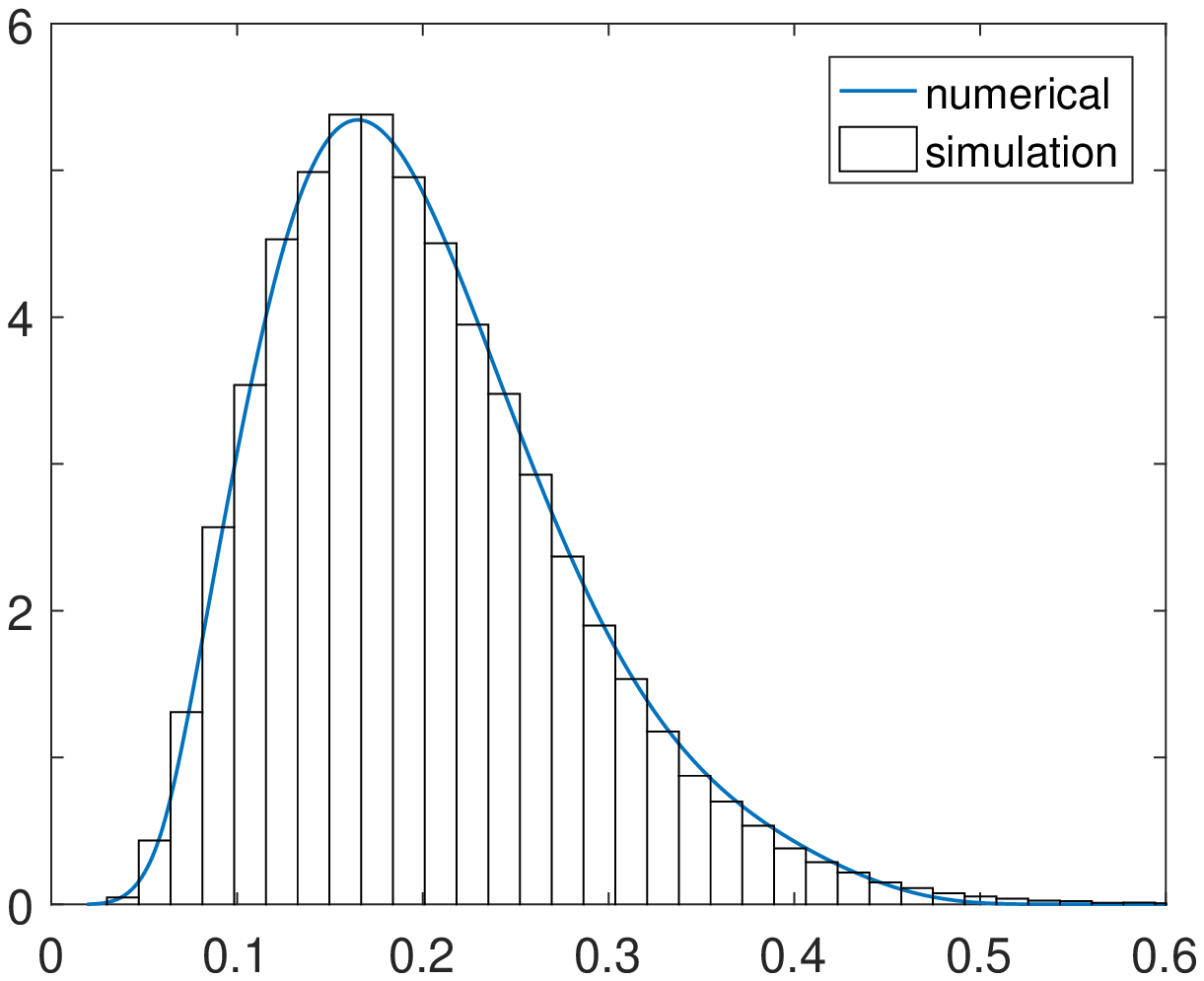}
\caption{$a = 0, b=1/2$}
\end{subfigure}
\begin{subfigure}[b]{0.45\textwidth}
\includegraphics[width=\textwidth]{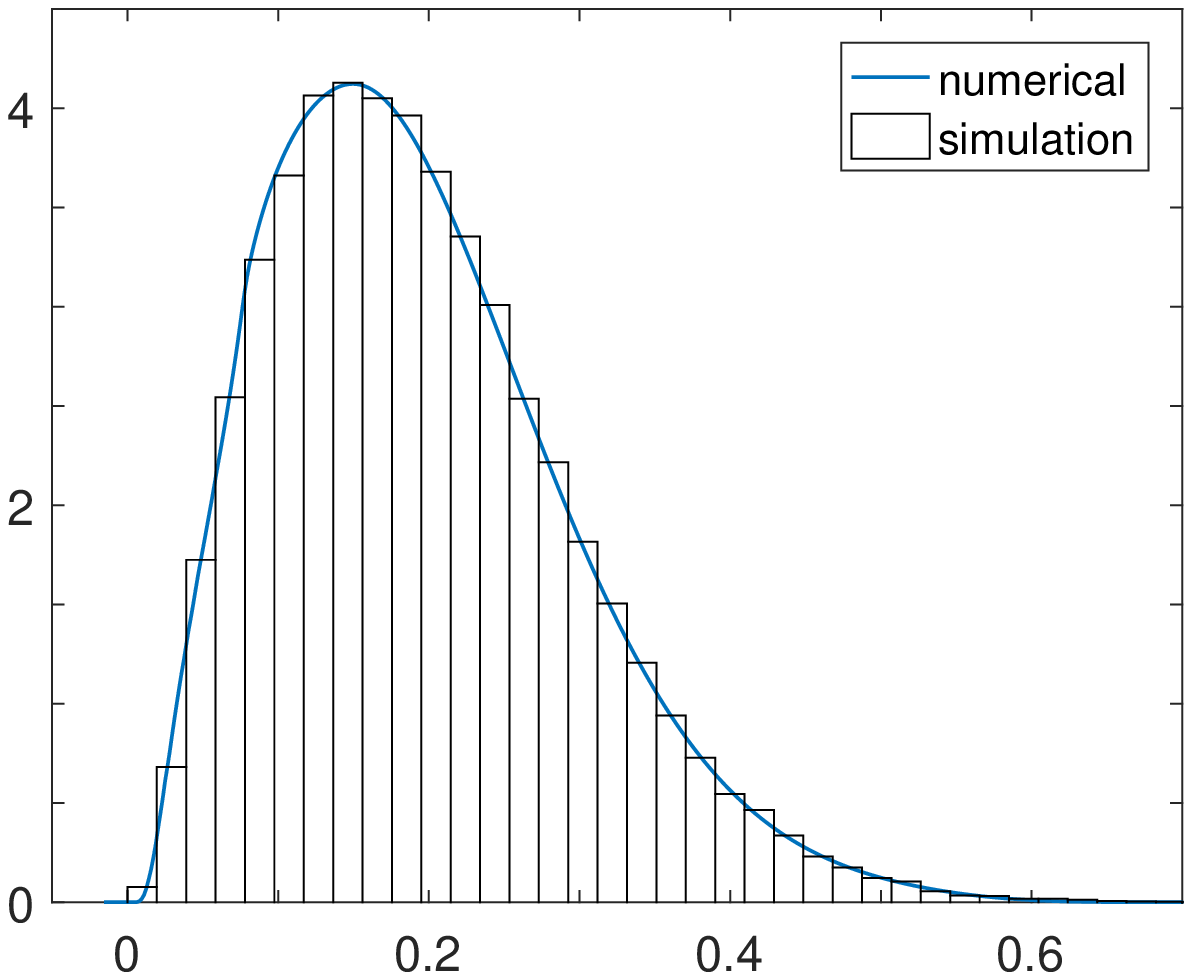}
\caption{$a = 1, b=1/2$}
\end{subfigure}
\\
\begin{subfigure}[b]{0.45\textwidth}
\includegraphics[width=\textwidth]{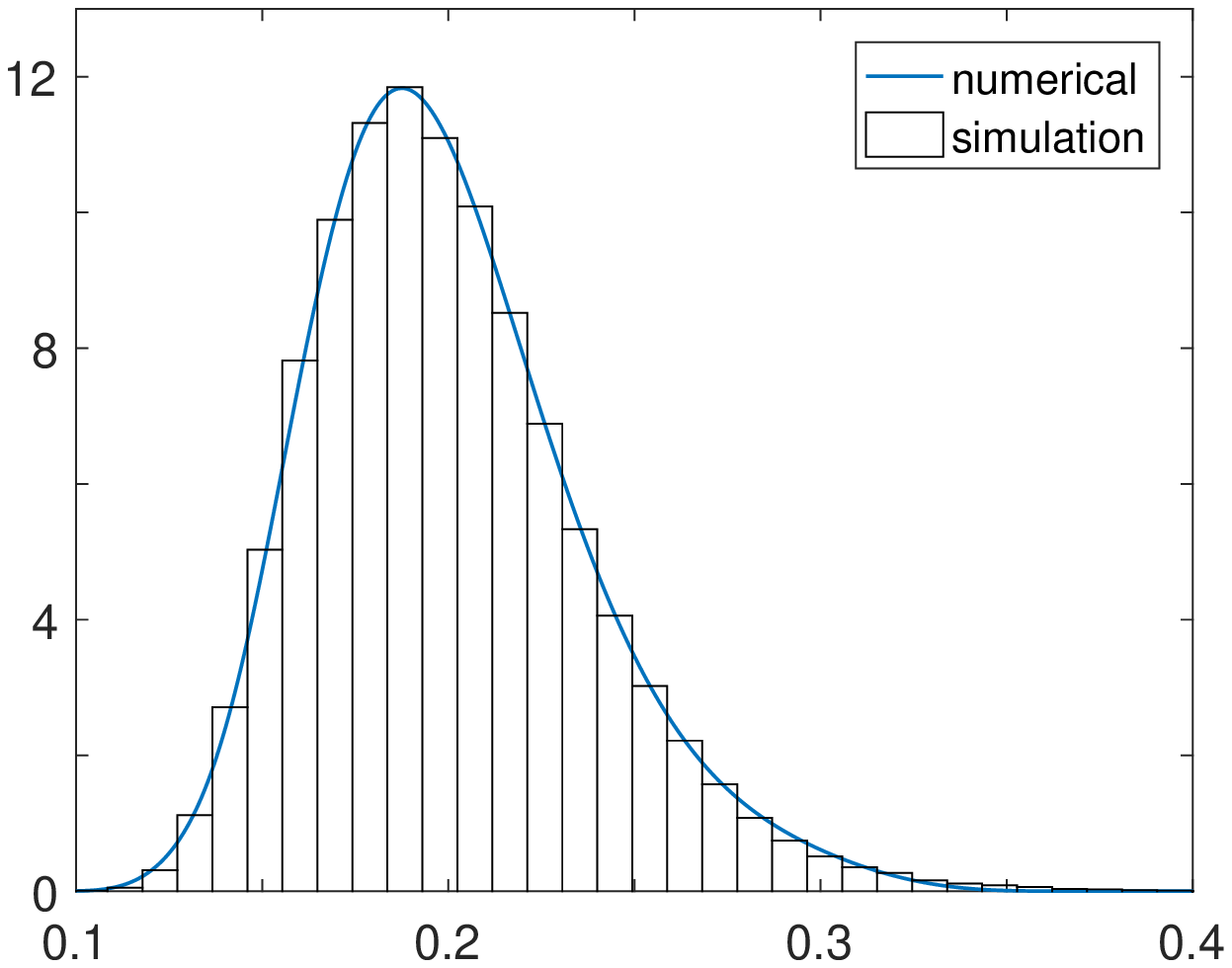}
\caption{$a = 0, b=1$}
\end{subfigure}
\begin{subfigure}[b]{0.45\textwidth}
\includegraphics[width=\textwidth]{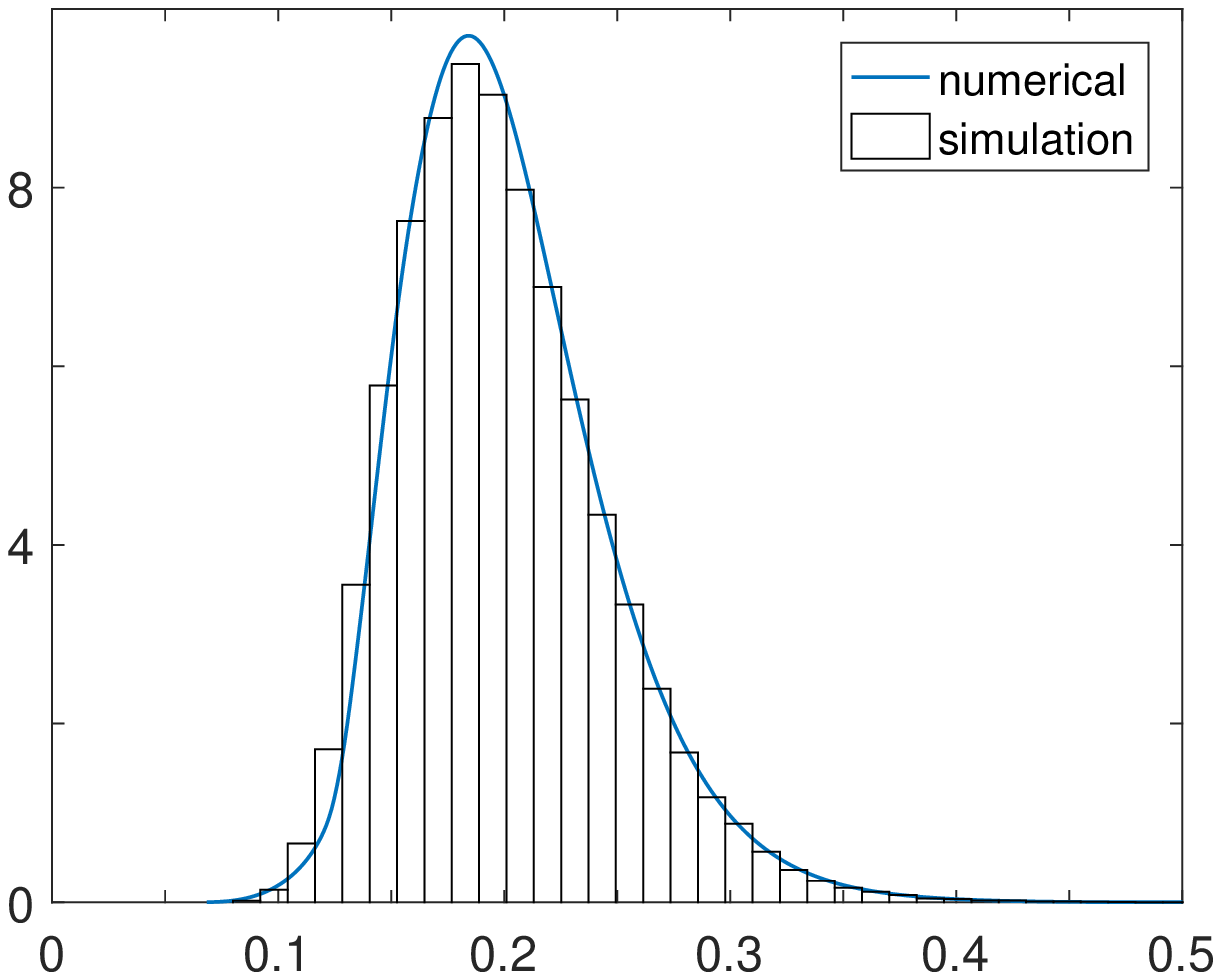}
\caption{$a = 1, b=1$}
\end{subfigure}
\\
\begin{subfigure}[b]{0.45\textwidth}
\includegraphics[width=\textwidth]{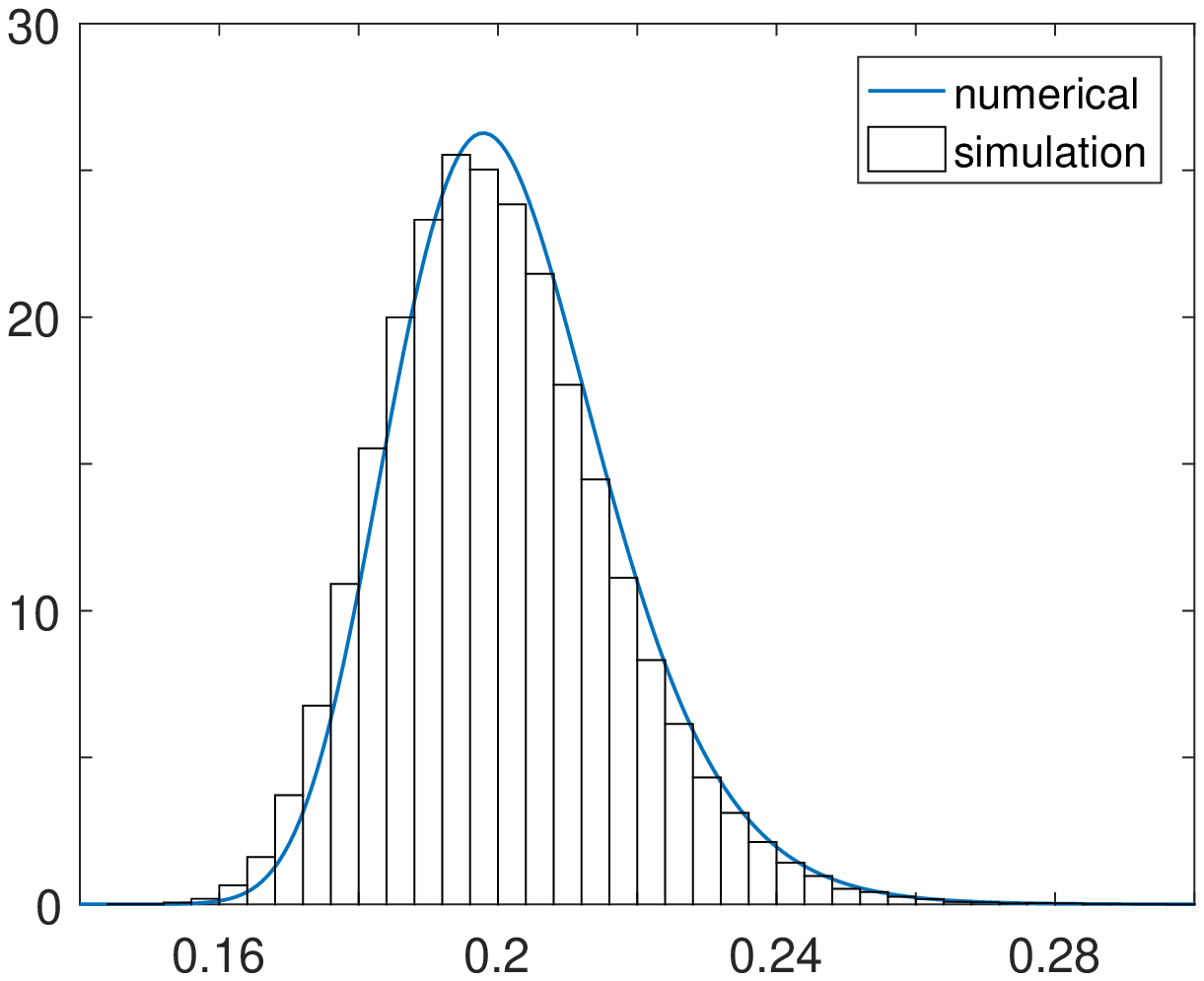}
\caption{$a = 0, b=3/2$}
\end{subfigure}
\begin{subfigure}[b]{0.45\textwidth}
\includegraphics[width=\textwidth]{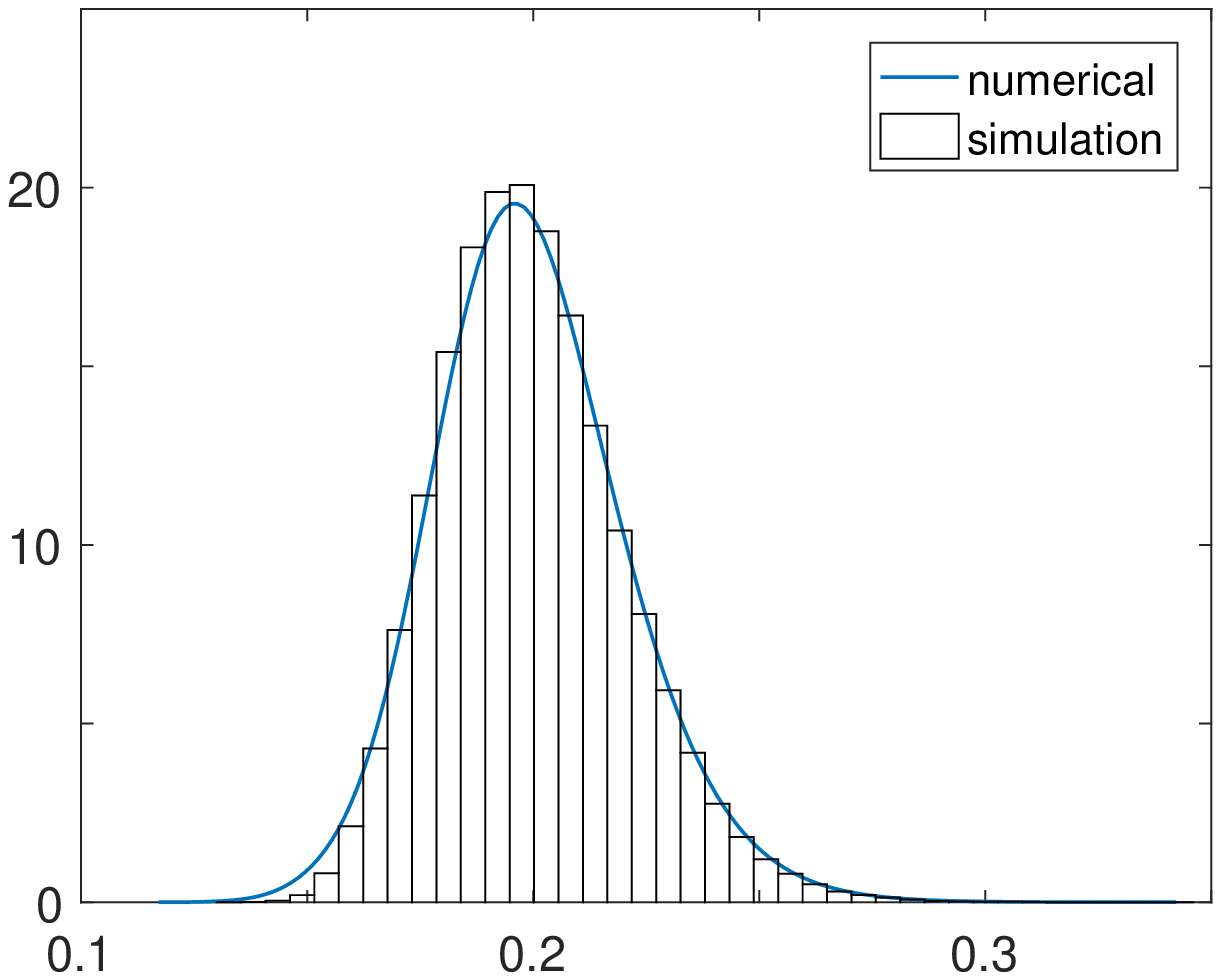}
\caption{$a = 1, b=3/2$}
\end{subfigure}
\caption{Probability density function of integrated variances $Y = IV_T$ with various stochastic volatility models}\label{Fig:IV}
\end{figure}

Using the recursive method with $h(x_n, x_{n+1}) = x_n$,
and approximating $$IV_t \approx \frac{1}{N} \sum_{n=1}^N V_n, $$
the probability density functions of the integrated variance under various stochastic models are presented in Figure~\ref{Fig:IV}.
The basic setting is the same as in previous cases, except for the formula of $h$.
Compared with the simulation results, the numerical density functions are precise for all stochastic volatility models.

\subsection{The GARCH model}\label{Subsect:GARCH}

We can calculate the numerical density function for not only continuous models but also discrete time models.
Consider the GARCH model \citep{bollerslev1986generalized} for the variance $\sigma^2$ and log-return $\varepsilon$:
$$ \sigma^2_{n+1} = \omega + \beta \sigma^2_n + \alpha \varepsilon^2_n, \quad \varepsilon_n \sim N(0, \sigma_n^2). $$
We are interested in the distributions of both time $N$ variance, $\sigma^2_N$, and total return, $\sum_{i=1}^{N} \varepsilon_i$.

The conditional distribution of $\alpha \varepsilon^2_n$ with given $\sigma_n$ is represented by the gamma distribution, such that
$$  \alpha \varepsilon^2_n | \sigma_n \sim \Gamma\left(\frac{1}{2}, 2\alpha\sigma_n^2 \right), $$
where the first argument of $\Gamma(,)$ is the shape parameter and the second argument is the scale parameter in the gamma distribution.
Therefore, the transition probability from $\sigma^2_{n}$ to $\sigma^2_{n+1}$
can be represented by the shifted gamma probability density function
such that
$$ f_{\sigma^2_{n+1} | \sigma^2_{n}}(x) = \frac{ \exp\left(-\frac{2\alpha\sigma_n^2}{x-\omega-\beta \sigma_n^2}\right)}{\sqrt{2\alpha\sigma_n^2(x-\omega-\beta\sigma_n^2)} \Gamma(1/2)}.$$

First, let us examine the distribution of $\sigma^2_{N}$ with given $\sigma_{0}$.
Since the closed form of the transition probability density $f_{\sigma^2_{n+1} | \sigma^2_{n}}$ is available, by setting $h(x_n, x_{n+1}) = x_{n+1} - x_{n}$, and hence, $ Y = \sigma^2_{N} - \sigma^2_{0}$, the theory is based on the recursive method.

\begin{figure}
\centering
\includegraphics[width=0.4\textwidth]{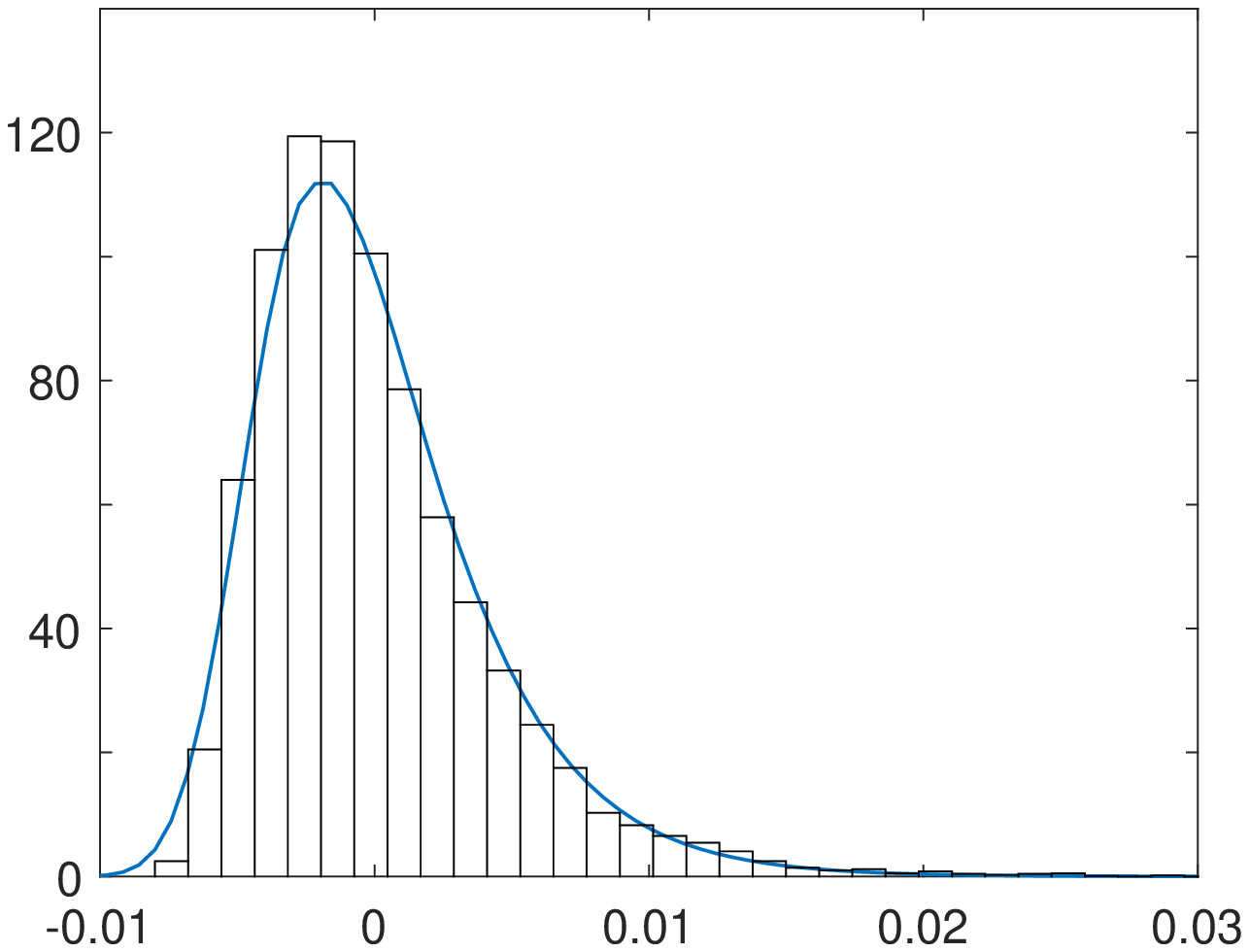}
\quad
\includegraphics[width=0.42\textwidth]{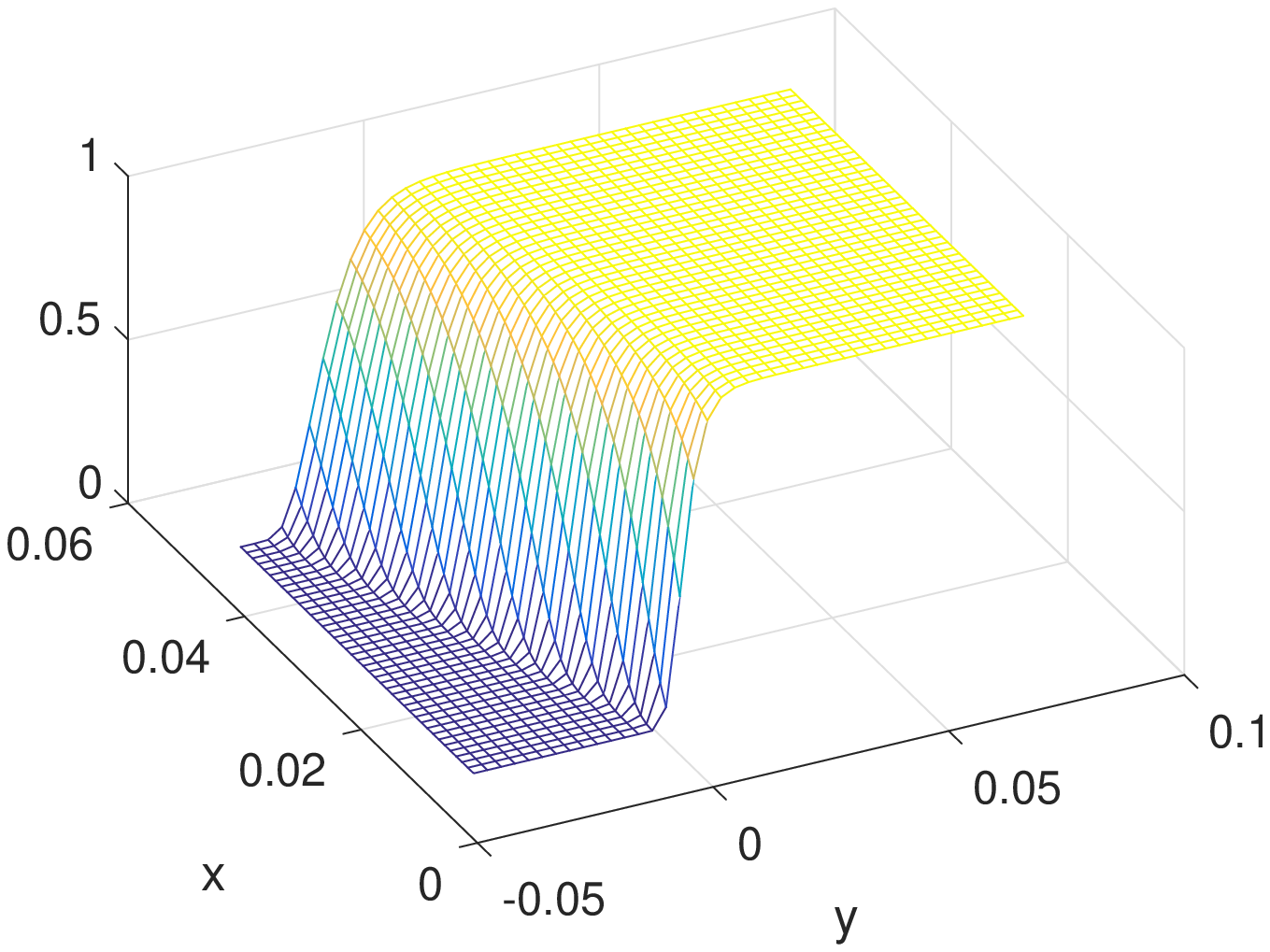}
\caption{Probability density function of the GARCH variance (left) and $F(y|x_0)$ (right)}\label{Fig:GARCH_V}
\end{figure}

Since the probability density function contains a non-smooth point, it is preferable to use the cumulative distribution function, $F_n$, for the recursive procedure.
Figure~\ref{Fig:GARCH_V} shows the histogram of the simulated GARCH process and numerically computed density function of $\sigma^2_{N}$ with $N = 20$.
For the parameter settings, we utilize $\omega = 0.001$, $\alpha = 0.05$ and $\beta = 0.9$.
For the numerical procedure, $x$-grid is set to be $[x_{\textrm{min}}, x_{\textrm{max}}] = [0.01, 0.05]$ with $\Delta x = 0.0002$
and the number of intervals for the $y$-axis is 150.
The tolerance level for $y$ is $10^{-8}$.
The right side of the figure presents the typical shape of $F_0$.
For any starting point $x_0$, along the $x$-axis, the conditional cumulative distribution function of $Y$ is retrieved along the $y$-axis.

\begin{figure}
\centering
\includegraphics[width=0.4\textwidth]{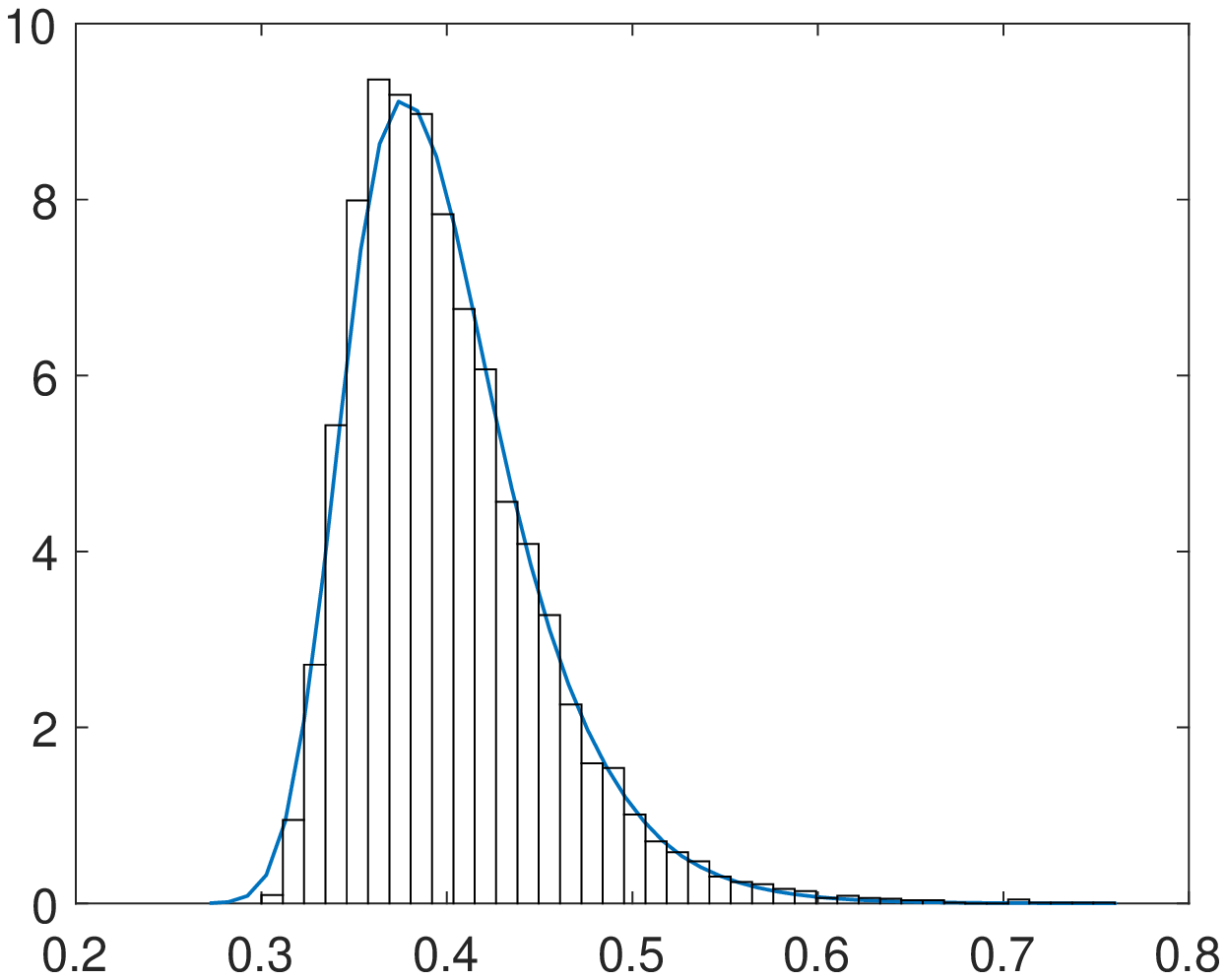}
\quad
\includegraphics[width=0.4\textwidth]{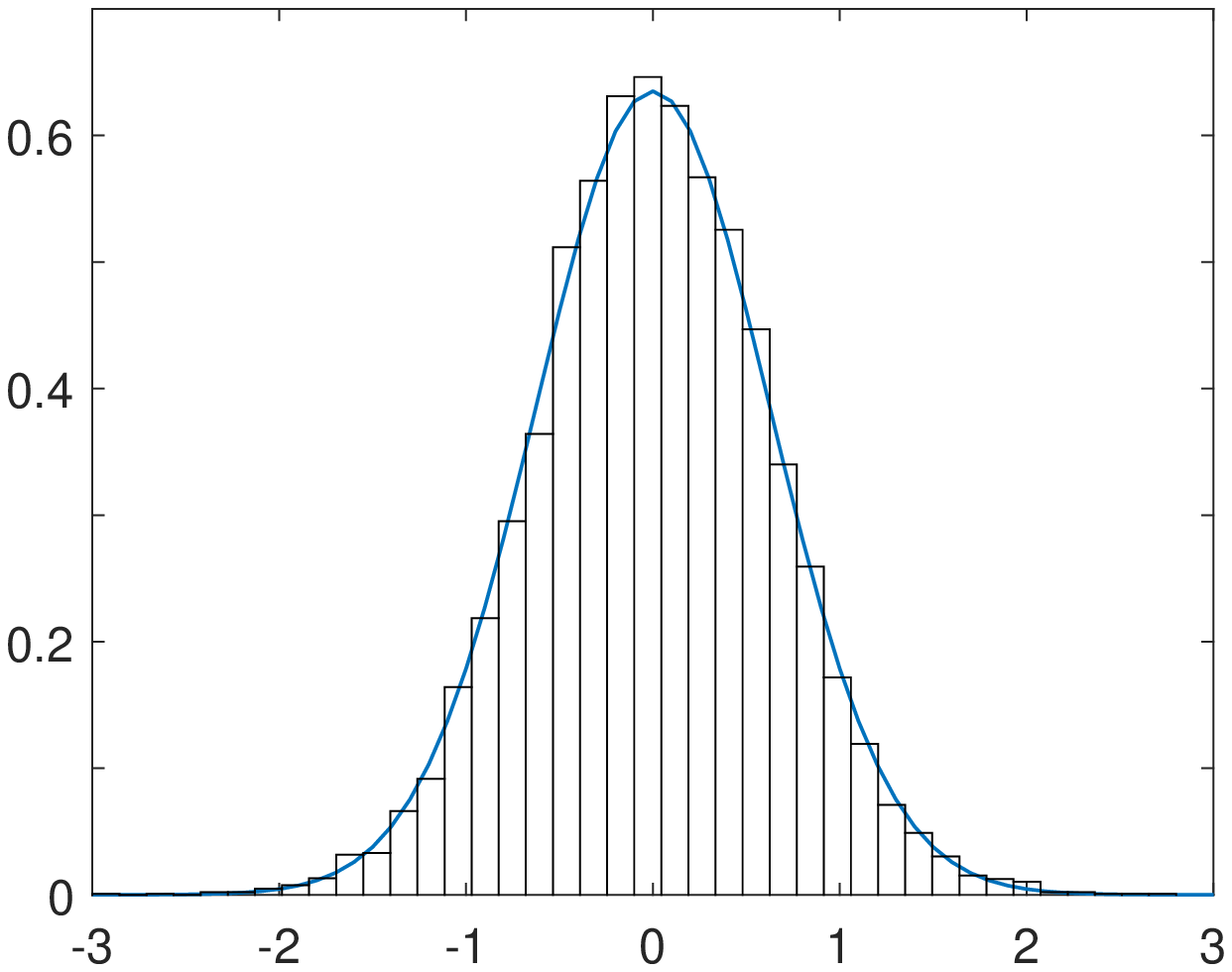}
\caption{Probability density function of the sum of GARCH variance (left) and return (right) }\label{Fig:GARCH_return}
\end{figure}

Second, we set $h(x_n, x_{n+1}) = x_n/N$, $Y = \sum_{i=1}^{N} \sigma^2_n/N$,
and compute $\sum_{i=1}^{N} \varepsilon_i$.
Since
$$\left. \sum_{i=1}^{N} \varepsilon_{i} \right| \frac{1}{N}\sum_{i=1}^{N} \sigma_n^2$$
follows the conditional normal distribution,
we can compute the total log-return from time $t_0$ to time $t_N$.
Figure~\ref{Fig:GARCH_return} presents the results.
On the left side is the probability density function of $\sum_{i=1}^{N} \sigma_n^2/N$ and the probability density function of $\sum_{i=1}^{N} \varepsilon_{i}$ is on the right side.
The figures show that the numerically computed GARCH return distribution is close to the simulation result.

\subsection{Hedging error}\label{Subsect:hedging}

Many researchers such as \cite{sepp2012approximate}, have studied errors occurring from hedging strategies of an option under time-discretization.
We consider hedging errors in a framework in which the underlying process follows the exponential L\'{e}vy model, as in \cite{madan1998variance}.
This subsection can be regarded as an extension of \cite{park2016distribution}.

Let $\gamma_t$ be a gamma process, which is a L\'{e}vy process with independent and gamma distributed increments,
with a mean rate parameter of 1 and variance parameter $\nu$.
In other words, the L\'{e}vy measure of $\gamma$ is represented by $ \nu z^{-1} \exp(- z/\nu )$ for jump size $z$.
Assume that the price of the underlying asset under a risk-neutral measure follows the exponential variance gamma model:
\begin{equation}
 S_t = S_0 \exp \left(rt + X_t +\omega t \right) \label{Eq:VG}
\end{equation}
where $X_t$ is a variance gamma process represented by a time changed Brownian motion
$$ X_t = \theta \gamma_t + \sigma W_{\gamma_t}$$
and $ \omega = (1/\nu) \log (1-\theta \nu - \frac{1}{2}\sigma^2 \nu )$.
The L\'{e}vy measure of the variance gamma process is represented by
$$ k(z) \D z = \frac{\exp\frac{\theta z}{\sigma^2}}{\nu |z|} \exp \left(-\frac{\sqrt{\frac{2}{\nu} + \frac{\theta^2}{\sigma^2}}}{\sigma}|z|\right)\D z. $$
\cite{madan1998variance} derived the probability density function of $S_t$, with $S_0=1$, as follows:
$$f(x) = \int_0^\infty \frac{1}{\sigma\sqrt{2 \pi g}} \exp \left(-\frac{(x-\theta g)^2}{2\sigma^2 g}\right) \frac{g^{\frac{t}{\nu}-1} \exp\left( - \frac{g}{\nu} \right) } { \nu^{\frac{t}{\nu}} \Gamma(t/\nu) }\D g. $$
From the above formula, the transition probability density function from $S(t_n) = x_n$ to $S(t_{n+1}) = x_{n+1}$ can be computed by numerical integration.

There are several practical methods to compute the European call option price under the variance gamma process.
We use the fast Fourier transform method based on the dampened option price as follows:
$$\e^{\alpha \log K} \E^{\mathbb Q} [\e^{-rT}(S_T - K)^+]$$
for some constant, $\alpha$, as explained in \cite{carr1999option}.
With this setting, the European call option price with $S_0 = 1$ and maturity, $T$, is represented by
$$\frac{\e^{-\alpha \log K}}{\pi} \int_0^\infty \e^{-\I v \log K} \frac{\e^{-r T}\psi_T(v-(\alpha+1)\I)}{\alpha^2 + \alpha - v^2 + \I(2\alpha + 1)v }\D v$$
where $\psi_T(u) = \E^\Q [\e^{\I u \log S_T}]$.

We test two ways of hedging the European call option $C$: delta hedging and minimal variance hedging, as proposed in \cite{follmer1986contributions}.
The minimal variance hedging ratio of the European call option, $C$, under the variance gamma model is represented by
$$ \phi_t = \frac{\frac{1}{S_{t-}} \int k(\D z) (\e^z-1)[C(t, S_{t-}\e^z) -C(t, S_{t-}) ] }{\int (\e^z-1)^2 k (\D z)}.$$
For a more detailed explanation, consult \cite{cont2007hedging}.

Consider an investor with a short position in the call option and hedging strategy with long position in the underlying asset.
Under the trading strategy $\phi$, the trading error between time $t_i$ and $t_{i+1}$ is defined by the difference between $t_{i+1}$-realized value of the hedged portfolio and the price of the risk-free asset as follows:
\begin{equation*}
\phi_{t_i} S_{t_{i+1}} - C_{t_{i+1}} - (1 + r \delta t)(\phi_{t_i} S_{t_i} - C_{t_i} ).
\end{equation*}
The total error is
\begin{equation}
\sum_{i=1}^{N}  \{ \phi_{i} S_{i+1} - C_{i+1} - (1 + r \delta t)(\phi_{i} S_{i} - C_i ) \}.\label{Eq:mv}
\end{equation}
Similarly, for the delta hedging strategy, the total error is
\begin{equation}
\sum_{i=1}^{N}  \{ D_{i} S_{i+1} - C_{i+1} - (1 + r \delta t)(D_{i} S_{i} - C_i ) \} \label{Eq:delta}
\end{equation}
where $D$ denotes the delta hedging ratio.
Since the terms inside the summation in Eqs.~\eqref{Eq:mv}~and~\eqref{Eq:delta} are functions of the underlying process, we can apply the numerical recursive method to compute the distribution of the hedging errors.

\begin{figure}
\includegraphics[width=0.45\textwidth]{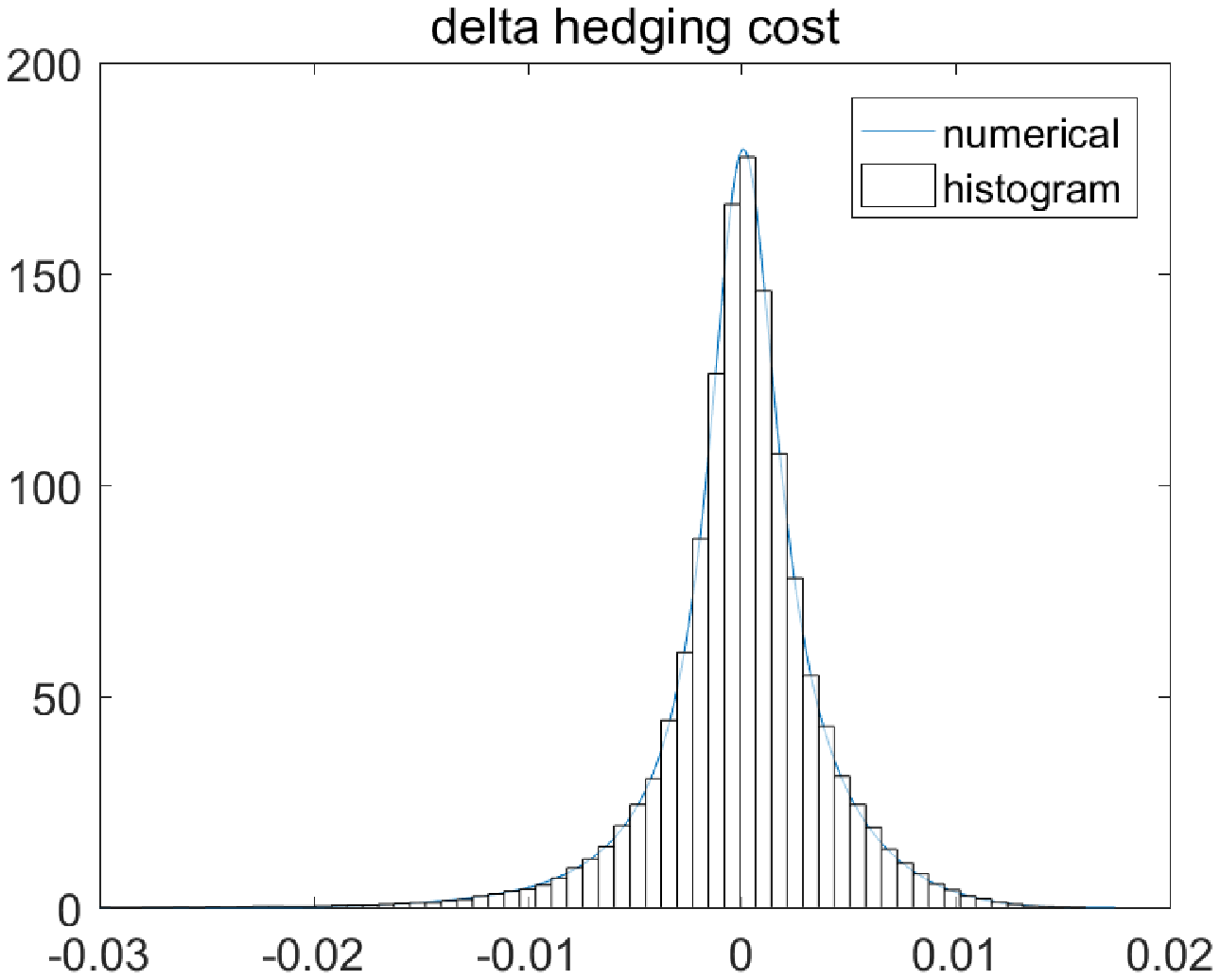}
\includegraphics[width=0.45\textwidth]{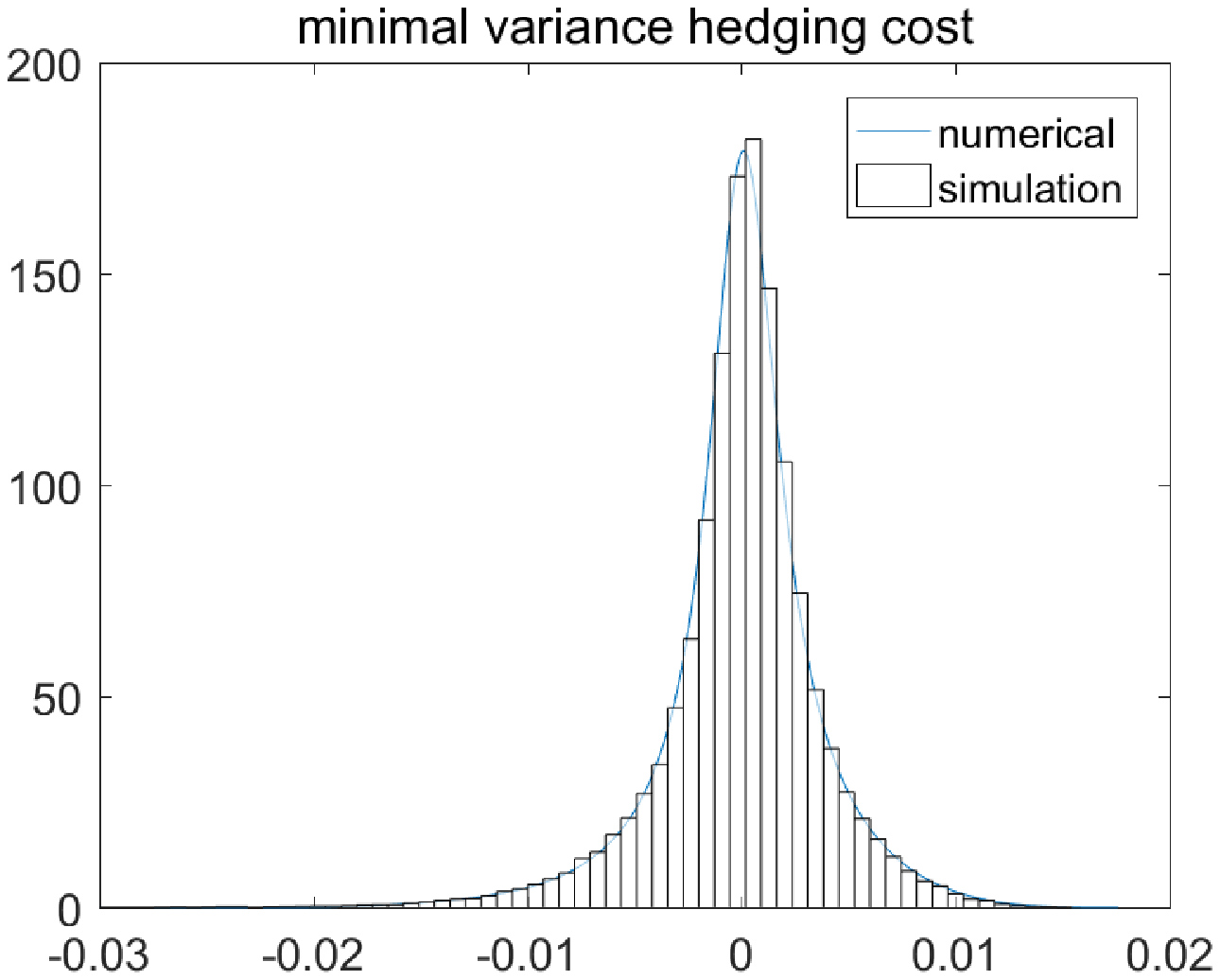}

\includegraphics[width=0.45\textwidth]{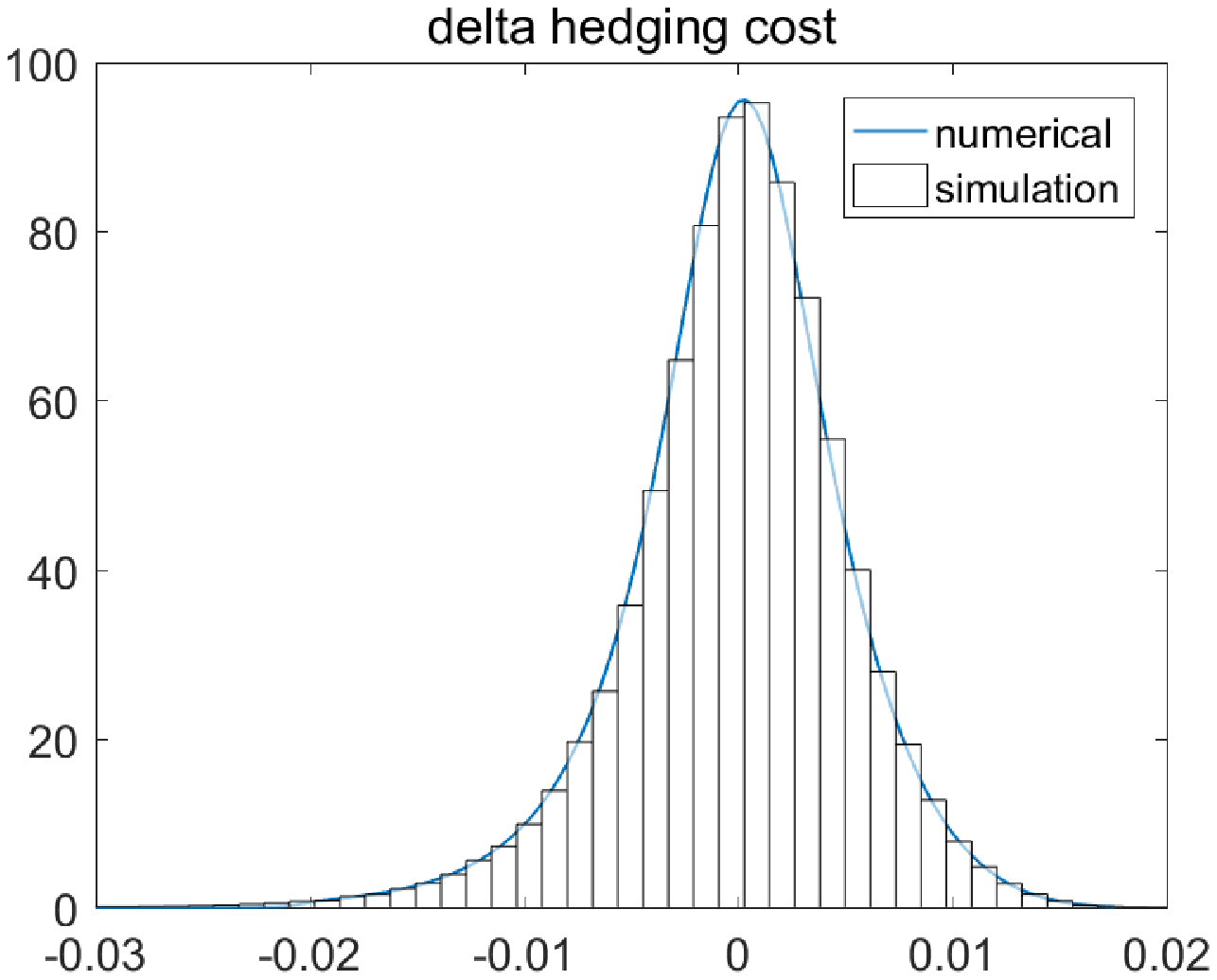}
\includegraphics[width=0.45\textwidth]{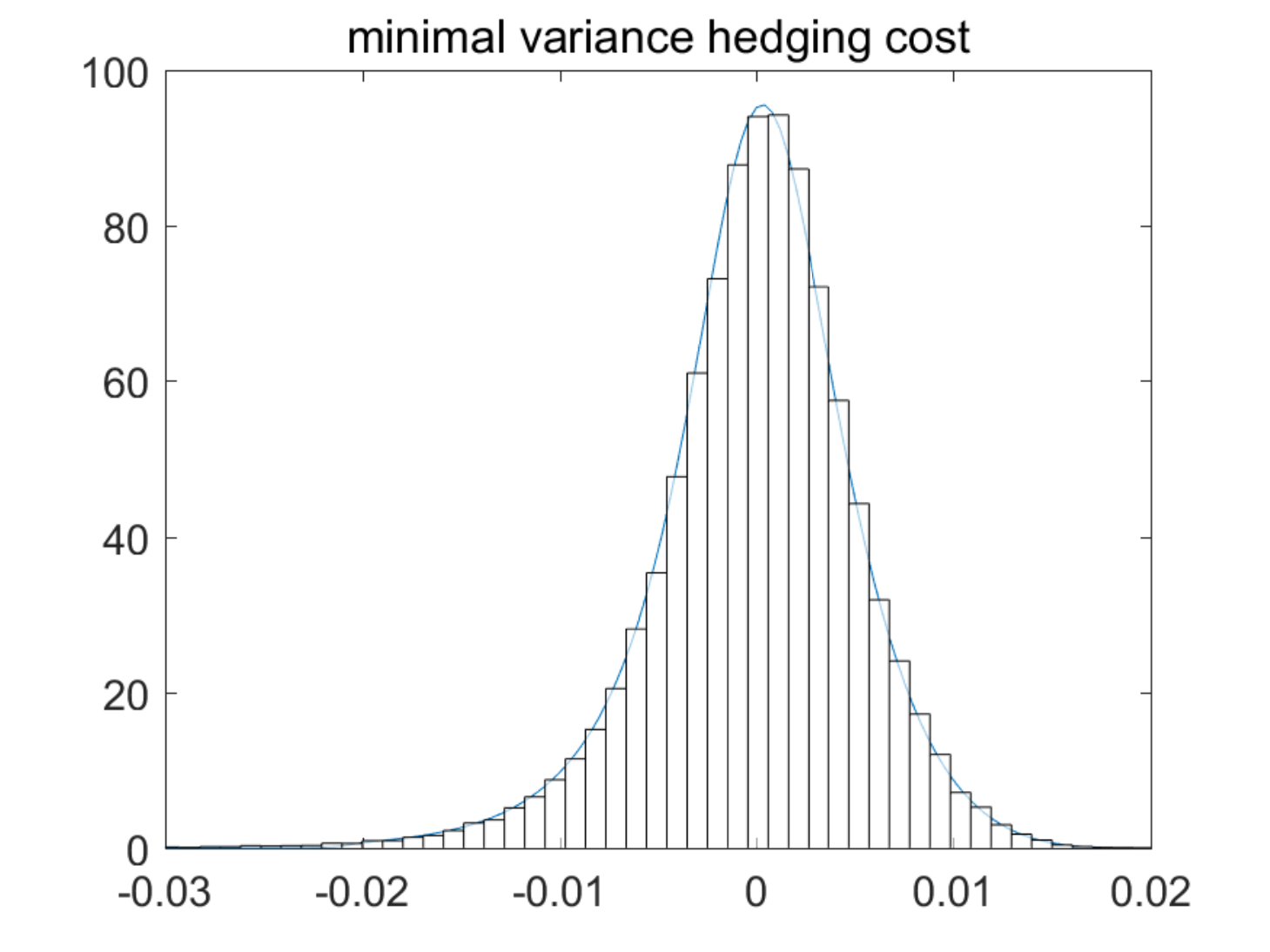}

\includegraphics[width=0.45\textwidth]{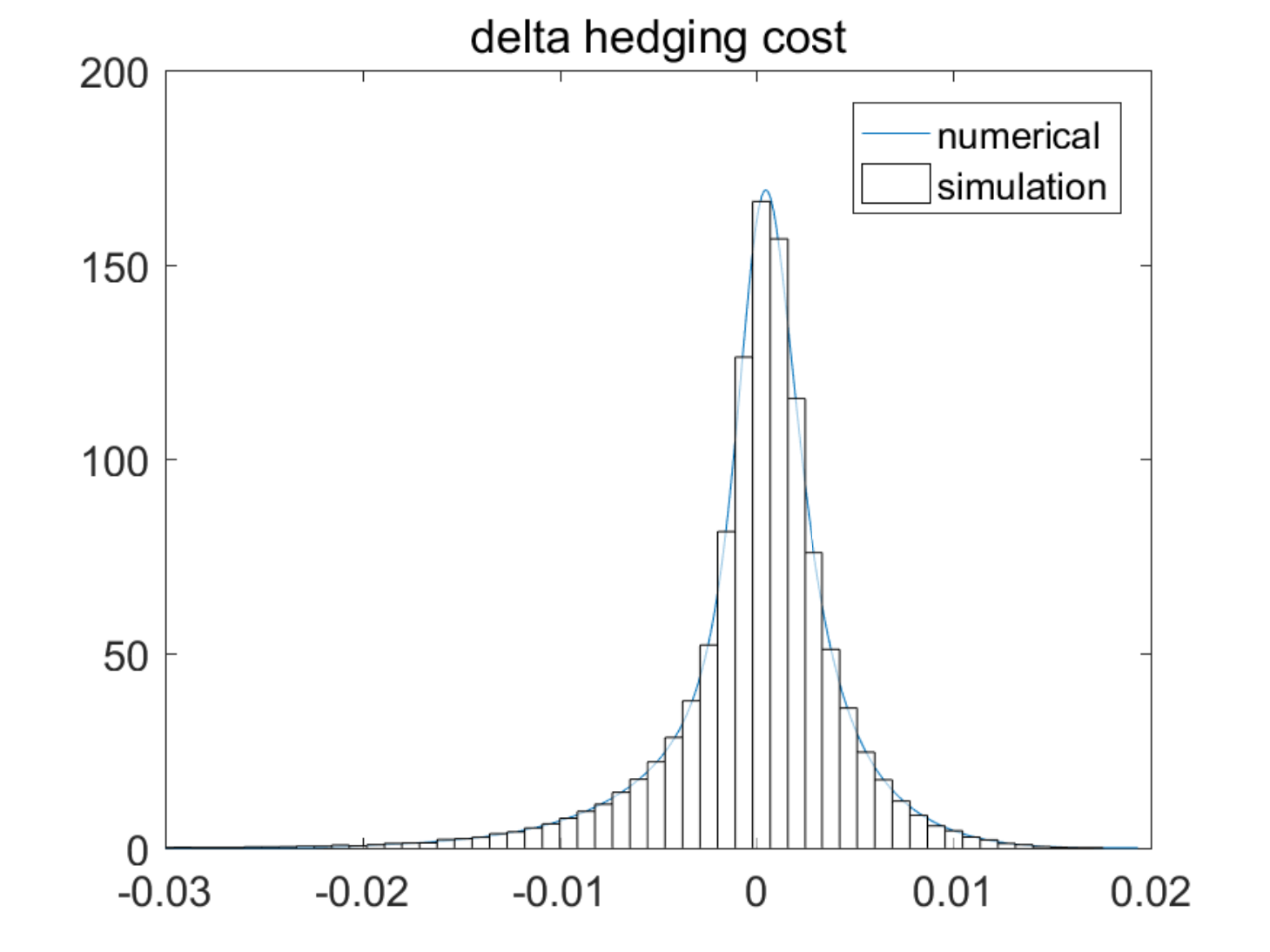}
\includegraphics[width=0.45\textwidth]{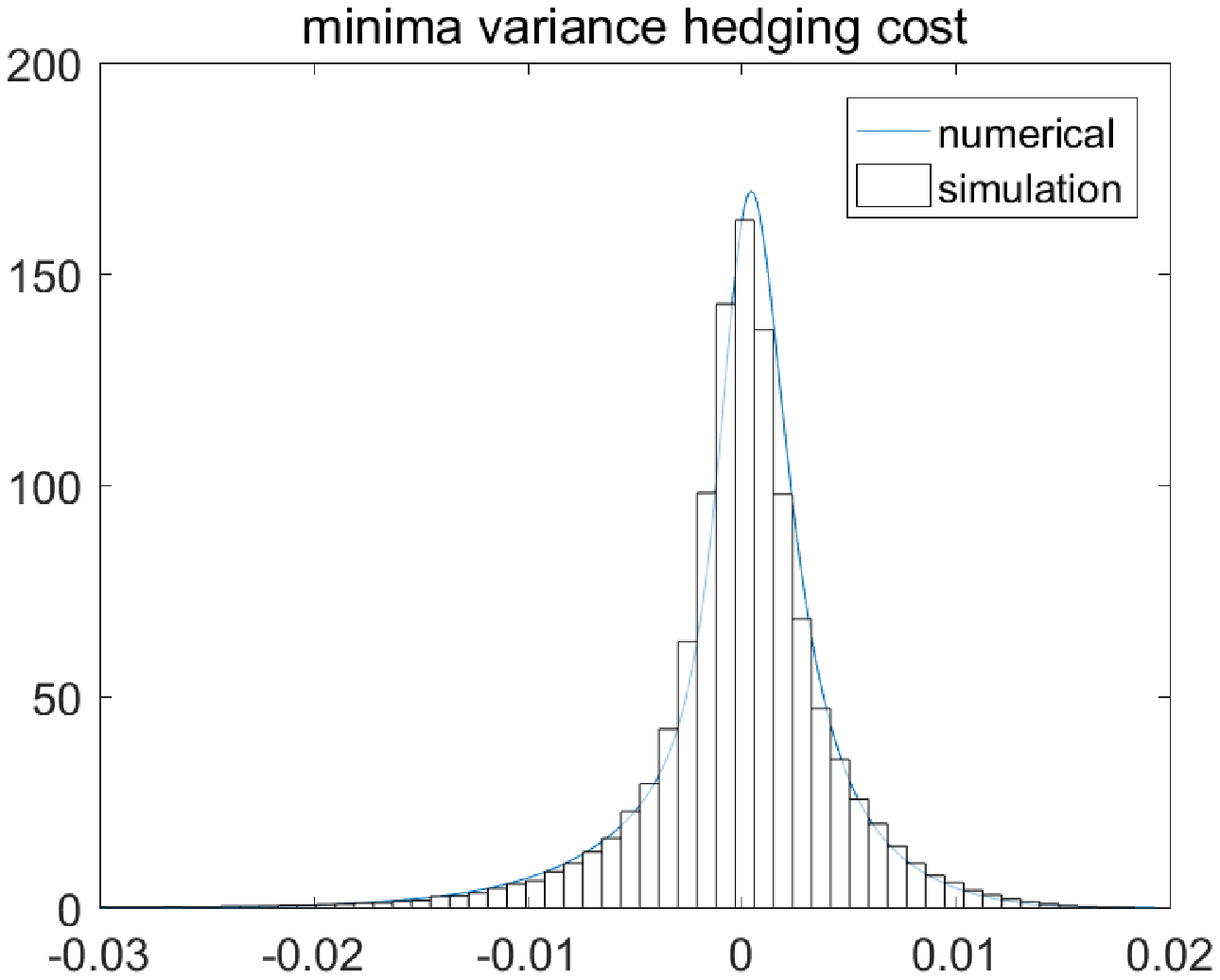}
\caption{Probability density functions of delta hedging errors (left) and minimum variance hedging errors (right) of European call options with strike $K = 0.9, 1, 1.05$ from top to bottom with $S_0 = 1$}\label{Fig:error}
\end{figure}

Figure~\ref{Fig:error} compares the numerically computed probability density functions of delta hedging errors (left) and minimum variance hedging errors (right) of European call options with the simulation histograms.
The European call options have strike price $K = 0.9, 1, 1.05$ with $S_0 = 1$ listed from top to bottom in Figure~\ref{Fig:error}.
The parameter setting is $\sigma = 0.2, \theta = 1.2, \nu = 0.001, r=0.02, \delta t = 1/250$.

\subsection{Arithmetic Asian option}

As mentioned, an Asian option is a financial derivative that is more robust to the manipulation of underlying asset prices than the European option.
Since the closed form formula for arithmetic Asian option prices is not known, simulation, approximation, or computational methods are generally used.
The recursive method can also be applied to compute the arithmetic Asian option prices.

Let $S$ be an underlying asset price process.
The arithmetic Asian option price over observation points $t_1 , \cdots, t_N = T$ with strike price $K$ is represented by
$$ \E^{\Q} \left[ \left( \frac{1}{N} \sum_{i=1}^N S_i  - K \right)^+ \right]$$
where the expectation is in terms of risk-neutral probability.
To compute the above expectation, we need to compute the distribution of $Y = \frac{1}{N} \sum_{i=1}^N S_i $.
Using the recursive method, setting $ h(x_{i-1}, x_i) = x_i/N$, or $ h(x_{i-1}, x_i) = x_i$, and subsequently, rescaling distribution,
we can compute the risk-neutral distribution of $Y$.
The Asian option price is determined using the recursive method whenever the risk-neutral transition probability is available.
The following can be considered an alternative method to that proposed in \cite{lee2014recursive}.

\begin{figure}
\centering
\includegraphics[width=0.65\textwidth]{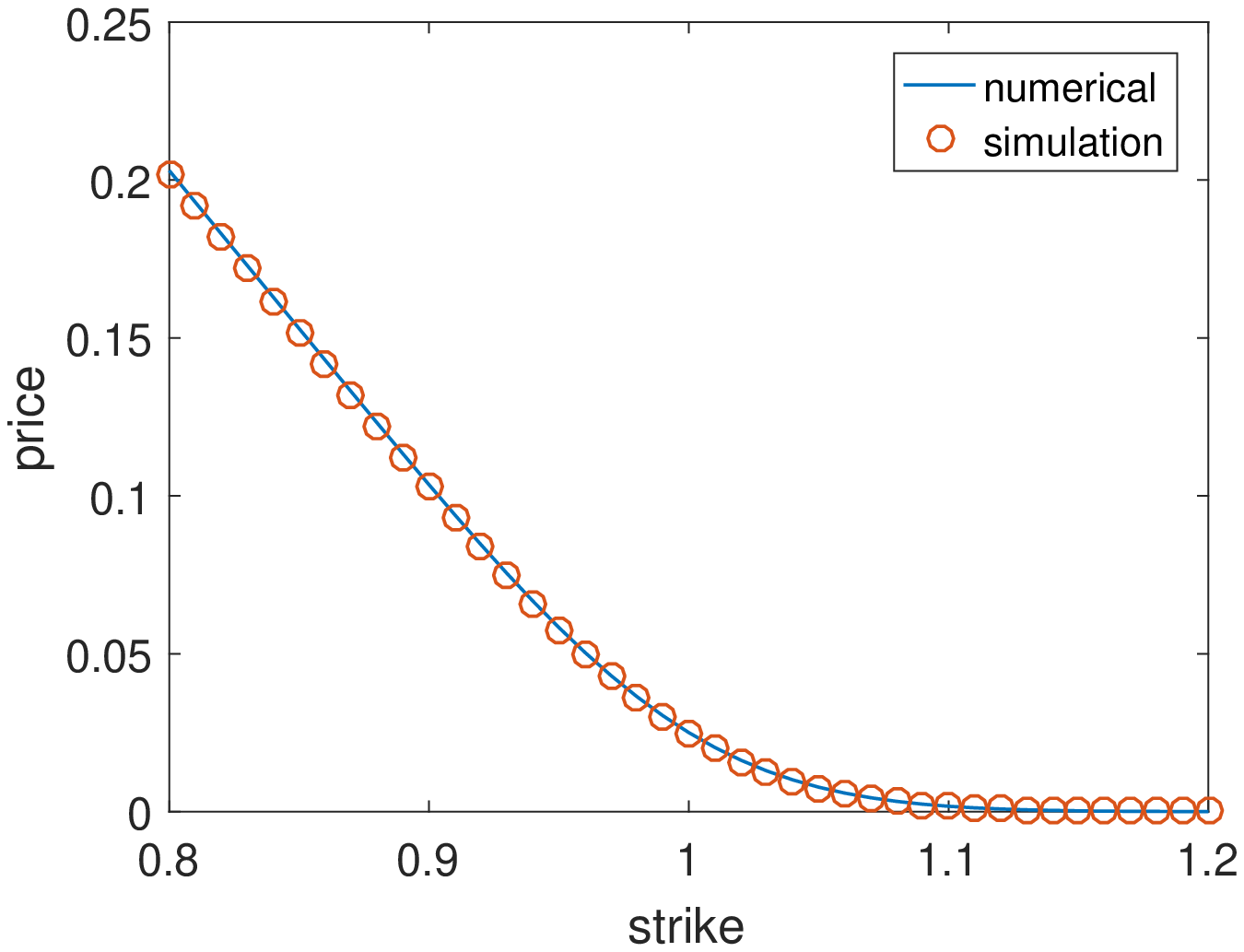}
\caption{Comparison between the Asian option price computed by numerical method and Monte Carlo simulation}
\label{Fig:asian}
\end{figure}

We assume that the stock price process follows a variance Gamma process, as in Eq~\eqref{Eq:VG}.
The parameter settings are $\sigma = 0.2, \theta = 1.2, \nu = 0.001, r=0.02$, and $S_0 = 1$.
The overall method is the same as in subsection~\ref{Subsect:hedging}, except the function form of $h$.
The comparison between the numerical method and simulation is presented in Figure~\ref{Fig:asian}
and the two results are quite similar.

\subsection{Skewness test}

In this section, we examine how the proposed recursive method can be used in hypothesis testing of the third moment of the asset return.
The distributions of asset returns tend to be skewed to the left.
Although it is generally not easy to measure the exact third moment of the return distribution,
the importance of the third moment has been acknowledged and extensively studied \citep{KrausLitzenberger, HarveySiddique}.

Let $R$ be a stationary return process.
For the hypothesis testing, the null hypothesis is
$ \mathrm{H}_0 : \E[ (\Delta R)^3] = 0 $ and the alternative hypothesis is $ \mathrm{H}_1 : \E[ (\Delta R)^3] < 0 $.
For example, consider a jump diffusion model as follows:
$$ \D R_t = \mu \D t + \sigma W_t + J \D N_t$$
where $\mu$ is drift, $\sigma$ is volatility, $N$ is a Poisson process with intensity $\lambda$, and $J$ follows a normal distribution with mean $\mu_J$ and standard deviation $\sigma_J$.
For simplicity, we set $\mu = -\lambda \mu_J$, such that $R$ becomes a martingale (with respect to suitable filtration).
Under this assumption, the statistical hypothesis can be modified as follows:
$\mathrm{H}_0 : \mu_J = 0$ and $\mathrm{H}_1 : \mu_J < 0$.
The test is similar to a simple t-test; however, the distribution of $(\Delta R)^3$ does not follow the normal distribution, and hence, it is advantageous to compute the exact distribution of $(\Delta R)^3$ using the recursive method.

\begin{figure}
\centering
\includegraphics[width=0.45\textwidth]{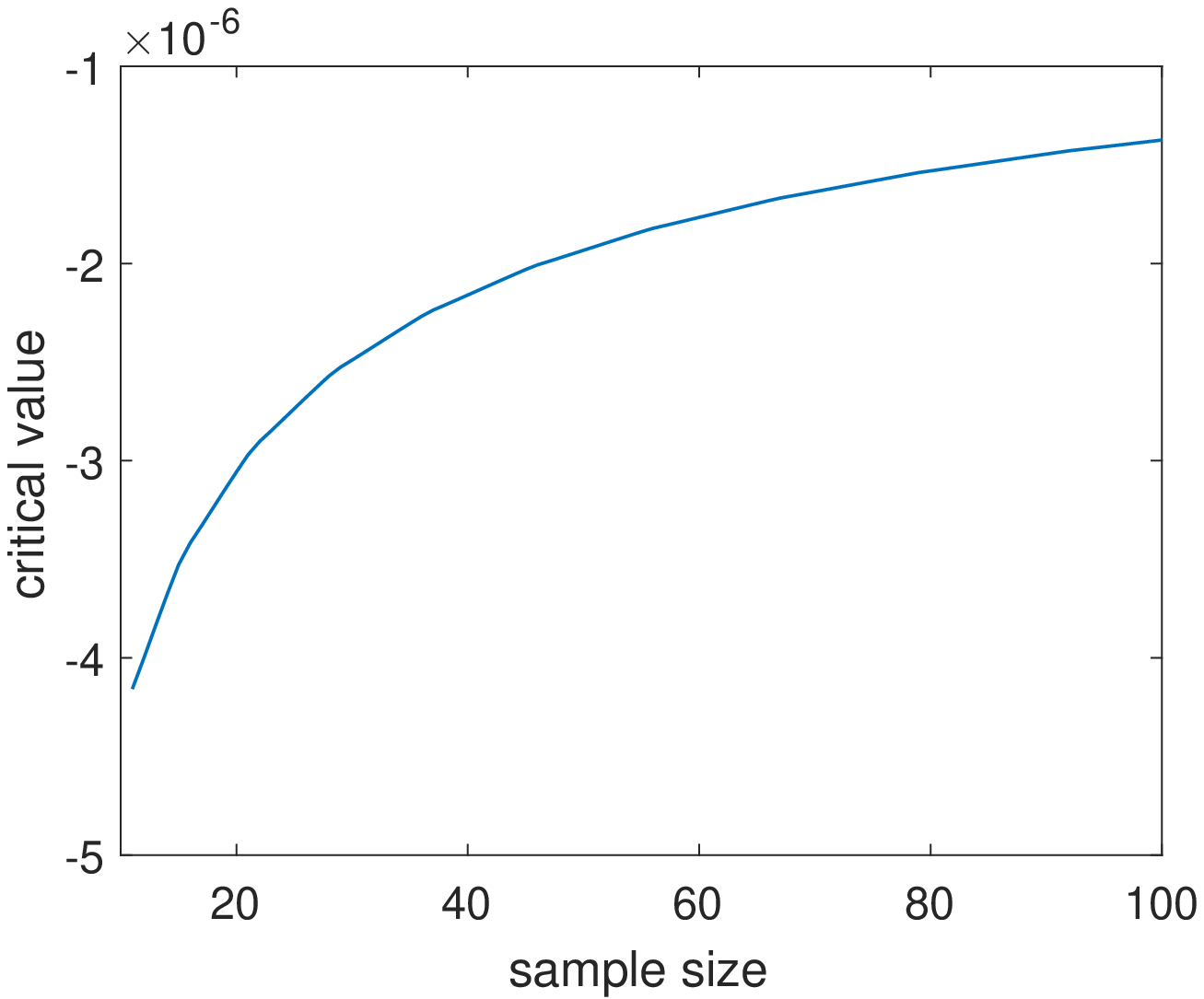}
\includegraphics[width=0.45\textwidth]{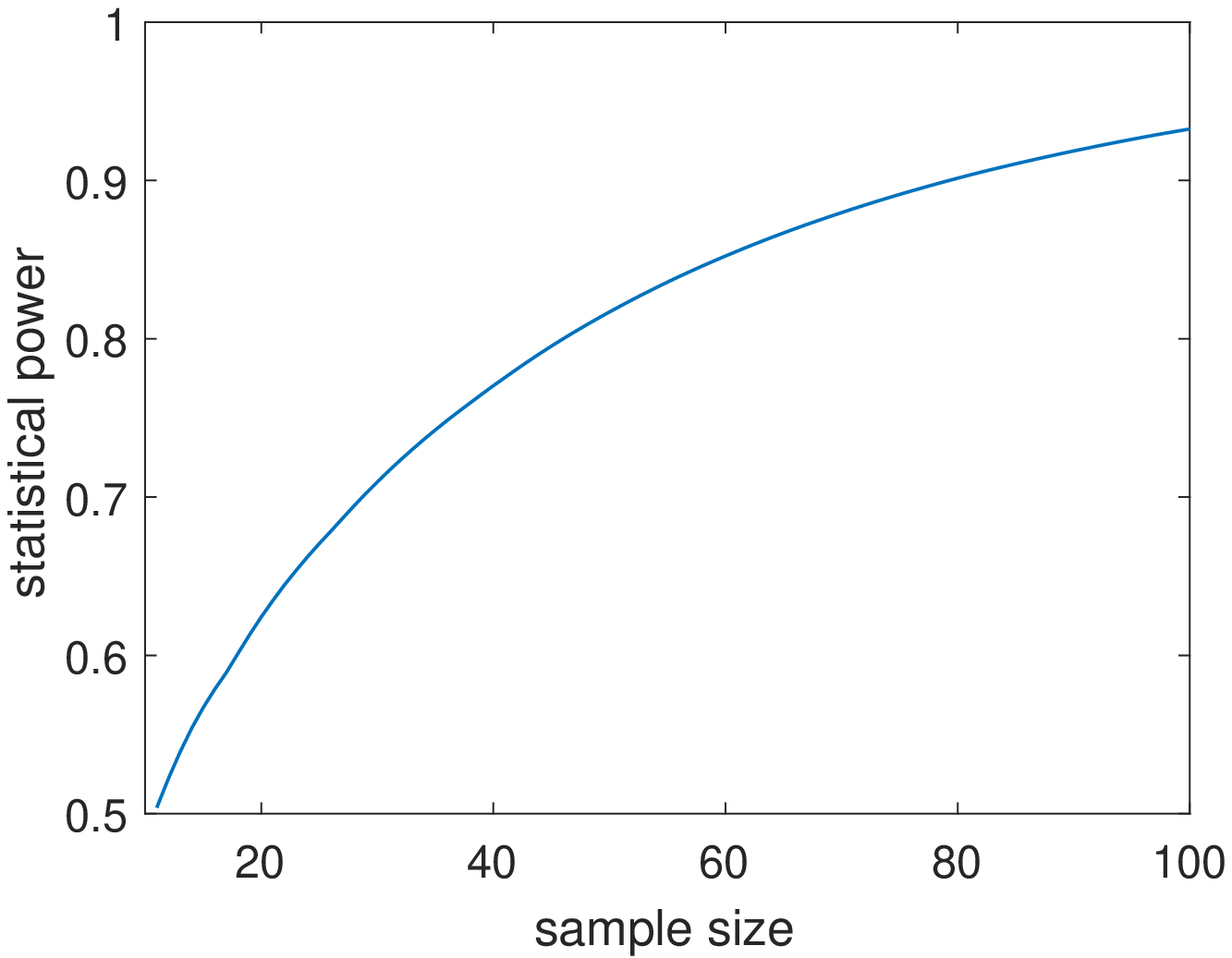}
\caption{Critical values for $(\Delta R)^3$ (left) and power curve (right) for testing $H_0 : E[(\Delta R)^3] = 0$ versus $H_1 : E[(\Delta R)^3] < 0$ with $\alpha = 0.05$}\label{Fig:Power}
\end{figure}

We compute the critical values that determine the rejection of the null hypothesis for sample sizes with given significance level, $\alpha = 0.05$ under the null hypothesis (see Figure~\ref{Fig:Power}).
The null hypothesis is rejected when the test statistic; the sample mean of $(\Delta R)^3$ is less than the corresponding critical value with given sample size.
To compute the critical values, the numerical probability density function of $(\Delta R)^3$ is computed to a
$h(x_n ,x_{n+1}) = (x_{n+1} - x_n)^3$ and the presumed parameter settings of $ \lambda = 10, \sigma_J = 0.01, \mu_J = 0$ and $\sigma = 0.1975 $.

The statistical power, typically denoted by $1-\beta$, is the probability that the test correctly rejects the null hypothesis when the null hypothesis is invalid.
We examine a power curve versus sample size
where $\mu_J$ is presumed to be $-0.05$, to imply negative skewness, and the other parameters are the same as in the previous case.
The right side of Figure~\ref{Fig:Power} shows the increase in the power curve with increasing sample size.
The curve implies that if we seek 90\% statistical power, this model will need approximately 90 samples.
This section presents an example of the third moments test, and the recursive method is deemed to be applicable for various statistical tests.

\section{Conclusion}\label{Sect:concl}

This study proposed a recursive formula for the distribution of specific functions and
detailed the application of the numerical procedure.
Various examples, including the numerical density function, the hedging error distribution, the arithmetic Asian option pricing, and statistical hypothesis testing, showed that the proposed method is quite precise.
The method is versatile, and we expect that more applications will become available not only in finance but also in various probabilistic analysis.
This study applied the method to a one-dimensional model, and
future studies can extend the method to the two-dimensional process model.

\bibliographystyle{apalike}

\newpage
\begin{thebibliography}{}
	
	\bibitem[Barndorff-Nielsen, 2002]{barndorff2002econometric}
	Barndorff-Nielsen, O.~E. (2002).
	\newblock Econometric analysis of realized volatility and its use in estimating
	stochastic volatility models.
	\newblock {\em Journal of the Royal Statistical Society: Series B (Statistical
		Methodology)}, 64:253--280.
	
	\bibitem[Bollerslev, 1986]{bollerslev1986generalized}
	Bollerslev, T. (1986).
	\newblock Generalized autoregressive conditional heteroskedasticity.
	\newblock {\em Journal of econometrics}, 31:307--327.
	
	\bibitem[Breeden and Litzenberger, 1978]{breeden1978prices}
	Breeden, D.~T. and Litzenberger, R.~H. (1978).
	\newblock Prices of state-contingent claims implicit in option prices.
	\newblock {\em Journal of business}, pages 621--651.
	
	\bibitem[Broadie and Kaya, 2006]{broadie2006exact}
	Broadie, M. and Kaya, {\"O}. (2006).
	\newblock Exact simulation of stochastic volatility and other affine jump
	diffusion processes.
	\newblock {\em Operations Research}, 54:217--231.
	
	\bibitem[Carr and Madan, 1999]{carr1999option}
	Carr, P. and Madan, D. (1999).
	\newblock Option valuation using the fast {F}ourier transform.
	\newblock {\em Journal of computational finance}, 2:61--73.
	
	\bibitem[Choe and Lee, 2014]{ChoeLee}
	Choe, G.~H. and Lee, K. (2014).
	\newblock High moment variations and their application.
	\newblock {\em Journal of Futures Markets}, 34:1040--1061.
	
	\bibitem[Christoffersen et~al., 2006]{Christoffersen2006}
	Christoffersen, P., Heston, S., and Jacobs, K. (2006).
	\newblock Option valuation with conditional skewness.
	\newblock {\em Journal of Econometrics}, 131:253--284.
	
	\bibitem[Christoffersen et~al., 2010]{Christoffersen2010}
	Christoffersen, P., Jacobs, K., and Mimouni, K. (2010).
	\newblock Volatility dynamics for the {S}\&{P}500: evidence from realized
	volatility, daily returns, and option prices.
	\newblock {\em Review of Financial Studies}, 23:3141--3189.
	
	\bibitem[Cont, 2001]{cont2001empirical}
	Cont, R. (2001).
	\newblock Empirical properties of asset returns: stylized facts and statistical
	issues.
	\newblock {\em Quantitative Finance}, 1:223--236.
	
	\bibitem[Cont et~al., 2007]{cont2007hedging}
	Cont, R., Tankov, P., and Voltchkova, E. (2007).
	\newblock Hedging with options in models with jumps.
	\newblock In {\em Stochastic analysis and applications}, pages 197--217.
	Springer.
	
	\bibitem[Cox et~al., 1985]{cox1985theory}
	Cox, J.~C., Ingersoll~Jr, J.~E., and Ross, S.~A. (1985).
	\newblock A theory of the term structure of interest rates.
	\newblock {\em Econometrica}, 53:385--407.
	
	\bibitem[Cox and Ross, 1976]{Cox1976}
	Cox, J.~C. and Ross, S.~A. (1976).
	\newblock The valuation of options for alternative stochastic processes.
	\newblock {\em Journal of financial economics}, 3:145--166.
	
	\bibitem[Fama, 1965]{fama1965behavior}
	Fama, E.~F. (1965).
	\newblock The behavior of stock-market prices.
	\newblock {\em The Journal of Business}, 38:34--105.
	
	\bibitem[F{\"o}llmer and Sondermann, 1986]{follmer1986contributions}
	F{\"o}llmer, H. and Sondermann, D. (1986).
	\newblock Hedging of non-redundant contingent claims.
	\newblock In {\em Contributions to Mathematical Economics: In Honor of
		G{\'e}rard Debreu}, pages 205--224. North Holland.
	
	\bibitem[French et~al., 1987]{french1987expected}
	French, K.~R., Schwert, G.~W., and Stambaugh, R.~F. (1987).
	\newblock Expected stock returns and volatility.
	\newblock {\em Journal of Financial Economics}, 19:3--29.
	
	\bibitem[Harvey and Siddique, 2000]{HarveySiddique}
	Harvey, C.~R. and Siddique, A. (2000).
	\newblock Conditional skewness in asset pricing tests.
	\newblock {\em Journal of Finance}, 55:1263--1295.
	
	\bibitem[Heston, 1993]{Heston1993}
	Heston, S.~L. (1993).
	\newblock A closed-form solution for options with stochastic volatility with
	applications to bond and currency options.
	\newblock {\em Review of Financial Studies}, 6:327--343.
	
	\bibitem[Kemna and Vorst, 1990]{kemna1990pricing}
	Kemna, A.~G. and Vorst, A. (1990).
	\newblock A pricing method for options based on average asset values.
	\newblock {\em Journal of Banking \& Finance}, 14:113--129.
	
	\bibitem[Kraus and Litzenberger, 1976]{KrausLitzenberger}
	Kraus, A. and Litzenberger, R.~H. (1976).
	\newblock Skewness preference and the valuation of risk assets.
	\newblock {\em Journal of Finance}, 31:1085--1100.
	
	\bibitem[Lee, 2014]{lee2014recursive}
	Lee, K. (2014).
	\newblock Recursive formula for arithmetic {Asian} option prices.
	\newblock {\em Journal of Futures Markets}, 34:220--234.
	
	\bibitem[Lee, 2016]{lee2016probabilistic}
	Lee, K. (2016).
	\newblock Probabilistic and statistical properties of moment variations and
	their use in inference and estimation based on high frequency return data.
	\newblock {\em Studies in Nonlinear Dynamics \& Econometrics}, 20:19--36.
	
	\bibitem[Madan et~al., 1998]{madan1998variance}
	Madan, D.~B., Carr, P.~P., and Chang, E.~C. (1998).
	\newblock The variance gamma process and option pricing.
	\newblock {\em European finance review}, 2:79--105.
	
	\bibitem[Musiela and Rutkowski, 2006]{musiela2006martingale}
	Musiela, M. and Rutkowski, M. (2006).
	\newblock {\em Martingale methods in financial modelling}, volume~36.
	\newblock Springer Science \& Business Media.
	
	\bibitem[Park et~al., 2016]{park2016distribution}
	Park, M., Lee, K., and Choe, G.~H. (2016).
	\newblock Distribution of discrete time delta-hedging error via a recursive
	relation.
	\newblock {\em East Asian Journal on Applied Mathematics}, 6:314--336.
	
	\bibitem[Sepp, 2012]{sepp2012approximate}
	Sepp, A. (2012).
	\newblock An approximate distribution of delta-hedging errors in a
	jump-diffusion model with discrete trading and transaction costs.
	\newblock {\em Quantitative Finance}, 12:1119--1141.
	
	\bibitem[V\v{e}c\v{e}r, 2002]{Vecer2002}
	V\v{e}c\v{e}r, J. (2002).
	\newblock Unified {Asian} pricing.
	\newblock {\em Risk}, 15:113--116.
	
\end{thebibliography}

\end{document}